\documentclass[journal ]{new-aiaa}

\usepackage[utf8]{inputenc}
\usepackage{textcomp}

\usepackage{graphicx}
\usepackage{amsmath}

\usepackage{booktabs}

\usepackage{amsthm}
\newtheorem{rem}{Remark}
\newtheorem{assum}{Assumption}
\newtheorem{define}{Definition}
\newtheorem{thm}{Theorem}
\newtheorem{lem}{Lemma} 
\usepackage[version=4]{mhchem}
\usepackage{siunitx}
\usepackage{longtable,tabularx}
\title{Control of Linear Dynamical Systems with \\ Operational Constraints}

\author{
	Marcel Menner\footnote{Boeing Designated Expert for Guidance/Autonomy, Senior Autonomy Engineer, Cambridge, MA 02474.},
	Eugene Lavretsky\footnote{Senior Principal Technical Fellow, Huntington Beach, CA 92647, AIAA Fellow.\\ \vspace*{0.5cm}\\ \hspace*{5cm}{{Copyright \copyright\ 2026. The Boeing Company. All Rights Reserved.}}}}
\affil{The Boeing Company}

\begin{document}

\maketitle  

\begin{abstract}
This paper presents a method for designing servo-controllers that enforce operational constraints on multi-input multi-output linear time-invariant dynamical systems. 
The proposed method introduces a control augmentation that complements a baseline controller to ensure forward invariance of state/output trajectories within prescribed box constraints. 
The derivations of the control design are based on the Nagumo Theorem, the Comparison Lemma, and principles from min-norm optimal control akin to the theory of Control Barrier Functions.
Importantly, an explicit piecewise-linear feedback control law is derived, which provides continuous enforcement of constraints without online optimization, enabling rigorous stability and robustness analysis via classical linear systems theory. 
Extensions to handle estimation errors in output feedback and integral action in proportional-integral control are provided. 
Theoretical results are validated through flight control simulation studies that demonstrate input/output constraint satisfaction, robustness to disturbances, and closed-loop stability. 
The proposed method offers a mathematically rigorous, practical, and analyzable alternative to common heuristic or optimization-intensive constrained control strategies. 
\end{abstract}

\section*{Nomenclature}

{\renewcommand\arraystretch{1.0}
	\noindent\begin{longtable*}{@{}l @{\quad=\quad} l@{}}
		$x$           & System state vector, $x \in \mathbb{R}^n$ \\
		$u$           & Control input vector, $u \in \mathbb{R}^m$ \\
		$y_{\lim}$    & Limited output vector subject to operational constraints \\
		$y_{\lim}^{\min}, y_{\lim}^{\max}$ & Component-wise minimum and maximum limits on $y_{\lim}$ \\
		$u_{\text{bl}}$ & Baseline control input vector \\
		$\pi$        & Control augmentation signal \\
		$A, B, C, D$ & State-space matrices of the system \\
		$C_{\lim}$    & Output matrix selecting the limited outputs from the state vector \\
		$C_{\lim,i}$  & $i$th row of matrix $C_{\lim}$ \\
		$H_{\pi}$    & Control sensitivity matrix relating control input to $r$-th derivative of constraint output \\
		$H_x$        & State sensitivity matrix relating state to $r$-th derivative of constraint output \\
		$\alpha_i, \alpha_{\pi}$ & Positive scalar or diagonal matrix gains tuning exponential convergence to constraint \\
		$r_i$        & Relative degree of the $i$th limited output (number of derivatives until control appears) \\
		$\lambda_{ij}$ & Real negative eigenvalues used to define stable polynomials in modified constraints \\
		$\Phi(s)$    & Diagonal polynomial matrix applying stable polynomials to constraints \\
		$H(x,u)$     & Vector of modified constraint inequalities after polynomial transformation \\
		$\Delta H_{\min}, \Delta H_{\max}$ & Offset vectors defining modified constraint residual terms \\
		$K$          & Feedback gain matrix of baseline controller \\
		$K_{i,\text{CBF}}$ & Feedback gain matrix for control barrier function (CBF) augmentation \\
		$F_i$        & Feedforward gain matrix for CBF commands related to constraint bounds \\ 
		$\delta(x)$  & Diagonal matrix of switching functions for all constraints \\
		$R_\pi$          & Positive definite weighting matrix used in quadratic cost of min-norm QP formulation \\
		$u_{\text{out}}$, $u_{\text{in}}$ & Signals for describing input-output relation in loop gain analysis \\
		$L_u(s; \delta)$ & Input breakpoint loop gain transfer function parameterized by active constraints $\delta$ \\
		$x_p$         & Plant state vector in PI controller extended system \\
		$y_{\text{reg}}$ & Regulated output vector in servo control context \\
		$e_y$        & Output tracking error, $e_y = y_{\text{reg}} - y_{\text{cmd}}$ \\
		$e_{y,I}$    & Integrated output tracking error including anti-windup augmentation \\
		$z_{\lim}$    & Limited outputs for PI controller design, subset or linear combinations of states \\
		$v$       & Anti-windup signal augmentation for integrator state \\
		$w$       & Proportional control augmentation signal \\
		$A_p, B_p, C_p, D_p$ & Plant subsystem matrices used in PI controller extension \\
		$K_I, K_P$   & Integral and proportional gain matrices in PI controller design \\
		$y$           & Measured output vector for output feedback case, $y \in \mathbb{R}^{n_y}$\\
		$\hat{x}$     & State estimate vector from observer  for output feedback case\\
		$e_{\text{est}}$ & State estimation error, $e_{\text{est}} = x - \hat{x}$ \\
		$L$          & Luenberger observer gain matrix in output feedback control \\ 
\end{longtable*}}

\section{Introduction}
\lettrine{L}{inear} system theory~\cite{Lavretsky2024,Hespanha2018} is vital in the design and analysis of control systems due to its rigorous mathematical foundation.
Specifically, robustness and control margins are critical aspects to ensure operational safety and practical functionality. 
While a true dynamical system is often nonlinear, controllers based on linear approximations often yield tractable and robust control solutions with prescribed specifications and operational requirements. 
Although linear approximations are generally valid within typical operating envelopes, their performance and robustness deteriorate when actuators saturate or outputs exceed predefined limits. 
In practice, operational constraints are often accommodated through heuristics or optimization-based approaches, which are challenging to analyze and may not be able to guarantee robustness.
This paper advances formal control methods by proposing a systematic approach to handle input/output operational boundaries while maintaining a continuous piece-wise linear state feedback control framework.
This ensures analyzability not only in typical operating conditions but also in the presence of operational constraints.

 \subsection{Related Work}

 Related methodologies for ensuring safety and constraint enforcement in control systems include Control Barrier Functions (CBFs) \cite{AmesXuGrizzleTabuada2016, WabersichZeilinger2022, WabersichEtAl2023} and Model Predictive Control (MPC) \cite{MorariLee1999}, both of which have seen increased interest for autonomous systems. MPC often functions as a supervisory "safety filter" that permits a baseline controller to operate unconstrained until system trajectories approach operational limits. CBFs commonly utilize scalar functions within an optimization framework to embed operational constraints and aim to achieve set invariance \cite{Blanchini1999, AmesCooganEgerstedtNotomistaSreenathTabuada2019}. Extensive research has extended CBFs to handle parametric uncertainties \cite{TaylorAmes2020}, higher relative degree constraints \cite{XiaoBelta2021, NguyenSreenath2016}, input delays \cite{MolnarKissAmesOrosz2022}, and real-time optimization \cite{AllibhoyCortes2023}. Applications employing these approaches range from automotive safety-critical systems \cite{SeoLeeBaekHorowitzChoi2022, AlanTaylorHeAmesOrosz2023} to robotics \cite{XiaoWangHasaniChahineAminiLiRus2023}. 
 
Distinct from these frameworks, this paper integrates vector-valued CBFs within a min-norm optimal control framework to derive an explicit analytical control law. 
The resulting piecewise-linear feedback controller activates only near operational boundaries and remains inactive otherwise. 
Unlike standard CBF or MPC methods, which typically require online optimization during operation, this approach employs classical control theory results such as the Nagumo Theorem \cite{Nagumo1942, MennerLavretsky2024}, the Comparison Lemma \cite{McNabb1986}, and min-norm optimal control principles \cite{FreemanKokotovic1996} only in the design phase, producing a closed-form solution based on the Karush-Kuhn-Tucker (KKT) conditions \cite{BoydVandenberghe2004}. This enables a rigorous stability and robustness analysis using classical control tools and facilitates certification for safety-critical applications, going beyond heuristic or numerical optimization-based strategies in related works. Related recent work focusing on analytical solutions to the CBF problem are discussed in~\cite{fisher2025analytical, cohen2026control}.

 \subsection{Contribution}
 
This work extends our initial investigations~\cite{LavretskyMenner2025, MennerLavretsky2026, menner2026output} by presenting a systematic approach for designing feedback controllers for multi-input multi-output (MIMO) linear time-invariant (LTI) systems subject to component-wise box constraint. 
It presents our collective works discussing state feedback~\cite{LavretskyMenner2025}, proportional-integral (PI) control~\cite{MennerLavretsky2026}, as well as output feedback~\cite{menner2026output}. 
The design framework integrates key theoretical tools with the Nagumo Theorem~\cite{Nagumo1942,MennerLavretsky2024}, the Comparison Lemma~\cite{McNabb1986}, and min-norm optimal control principles~\cite{FreemanKokotovic1996} to guarantee forward invariance of the closed-loop trajectories within prescribed operational limits on outputs of arbitrary relative degree.
These constraint conditions are embedded into a Quadratic Program (QP), which, with a carefully chosen cost function weighting, admits a closed-form analytical solution leveraging KKT conditions~\cite{BoydVandenberghe2004}.

 The resulting continuous piecewise-linear feedback control law augments a baseline linear servo-controller; it activates only as system trajectories approach operational boundaries and remains inactive otherwise.
 Beyond enforcing "soft" box constraints, distinguished from traditional "hard" saturations implemented via static clipping, the augmentation offers a rigorous and systematic approach for operating in the presence of constraints.
 It also provides systematic and analyzable anti-windup mechanism for integral controllers, thus avoiding heuristic or ad hoc bounding strategies~\cite{AstromMurray2021}, i.e., by construction, the augmentation prevents the integrator from continuing to accumulate error due to actuators being saturated.
 The explicit analytical form facilitates rigorous analyses of closed-loop stability and robustness margins, including for observer-based output feedback and PI control. 
 The practical utility of our approach is demonstrated through flight control case studies showing effective constraint management, bounded tracking errors, and robust closed-loop stability despite nonlinearities and disturbances.
 
 \section{Problem Formulation}
 
 Consider the stabilizable MIMO LTI dynamical system,
 \begin{equation}
 	\label{eq1}
 	\begin{aligned}
 		\begin{matrix}
 			\dot{x}\left( t \right)=Ax\left( t \right)+Bu\left( t \right) \\ 
 			{{y}_{\lim }}\left( t \right)={{C}_{\lim }}\,x\left( t \right), 
 		\end{matrix}
 	\end{aligned}
 \end{equation}
 where $x\left( t \right)\in {{R}^{n}}$ is the state vector, $u\left( t \right)\in {{R}^{m}}$ is the control inputs, and ${{y}_{\text{lim}}}\left( t \right)\in {{R}^{m}}$ is the vector of limited outputs. This paper proposes a linear control design in presence of operational constraints on the limited outputs ${{y}_{\lim }}\left( t \right)$, 
 \begin{equation}
 	\label{eq2}
 	\begin{aligned}
 		y_{\lim }^{\min }\le {{y}_{\lim }}\left( t \right)\le y_{\lim }^{\max }\text{  }\Leftrightarrow \text{  }\begin{matrix}
 			{{h}_{\min }}\left( x\left( t \right) \right)=y_{\lim }^{\min }-{{C}_{\lim }}\,x\left( t \right)\le 0  \\
 			{{h}_{\max }}\left( x\left( t \right) \right)={{C}_{\lim }}\,x\left( t \right)-y_{\lim }^{\max }\le 0  \\
 		\end{matrix}
 	\end{aligned}
 \end{equation}
 with the component-wise minimum and maximum limits (box constraints), $y_{\lim }^{\min }$ and $y_{\lim }^{\max }$.
 Consider the control law,
 \begin{equation}
 	\label{eq3}
 	\begin{aligned}
 		u\left( t \right)={{u}_{\text{bl}}}\left( t \right)+\pi \left( t \right)
 	\end{aligned}
 \end{equation}
 with a control augmentation signal $\pi(t)$ and the baseline controller $u_{\rm bl} (t)$. The main goal of the methods in this paper is to derive a piecewise-linear control augmentation signal $\pi(t)$ that becomes active only if needed to avoid violating the operational constraints in~\eqref{eq2}.

\begin{define}[Forward Invariance \cite{Nagumo1942,MennerLavretsky2024}]
	For the system $\dot{x}=f(x)\in {{R}^{n}}$	defined in $\mathcal{D}\subset {{R}^{n}}$, a set $\mathcal{E}\subset \mathcal{D}$ is called forward invariant in $\mathcal{D}$ if for every $x({{t}_{0}})\in \mathcal{E}$ with initial time ${{t}_{0}}$, $x(t)\in \mathcal{E}$, for all $t\ge {{t}_{0}}.$
\end{define}

\begin{define}[CBF-ability]
	\label{def1} 
	Let $r_i$ be the relative degree of the $i$-th limited output, i.e., $(y_{\rm lim} )_i$ in~\eqref{eq1} has to be differentiated $r_i$ times until the CBF augmentation $\pi(t)$ appears explicitly, see Section~\ref{sec3} for details. Further, let 
	\begin{equation}
		\label{eq4}
		\begin{aligned}
			{{H}_{x}}=\left[ \begin{matrix}
				{{\left( {{C}_{\lim }} \right)}_{1}}\prod\limits_{j=1}^{{{r}_{1}}}{\left( A-{{\lambda }_{1\,j}}{{I}_{n}} \right)}  \\
				\vdots   \\
				{{\left( {{C}_{\lim }} \right)}_{m}}\prod\limits_{j=1}^{{{r}_{m}}}{\left( A-{{\lambda }_{m\,j}}{{I}_{n}} \right)}  \\
			\end{matrix} \right],\text{    }{{H}_{\pi }}=\left[ \begin{matrix}
				{{\left( {{C}_{\lim }} \right)}_{1}}{{A}^{{{r}_{1}}-1}}  \\
				\vdots   \\
				{{\left( {{C}_{\lim }} \right)}_{m}}{{A}^{{{r}_{m}}-1}}  \\
			\end{matrix} \right]B,
		\end{aligned}
	\end{equation}
	where $(C_{\rm lim} )_i$ is the $i$th row of $C_{\rm lim}$. Then, system~\eqref{eq1} is said to be CBF-able if $H_\pi$ is nonsingular and there exist negative real eigenvalues $\lambda_{ij}<0$ such that $A-BH_\pi^{-1} H_x$ is Hurwitz.
\end{define}

 \begin{assum}
 	\label{assum1}
 	The matrix pair $\left( A,\,B \right)$ is stabilizable.
 \end{assum}
 \begin{assum}
 	\label{assum2}
 The dynamical system~\eqref{eq1} has a well-defined vector relative degree, see Section~\ref{ssec3a} for details.
 \end{assum}
 \begin{assum}
 	\label{assum3}
The dynamical system~\eqref{eq1} is CBF-able.
 \end{assum}

 \section{Generic Augmentation Design: State Feedback Control}
 \label{sec3}
 This section focuses on the generic state feedback case and derives a piecewise-affine control augmentation policy of the form 
 \begin{equation}
 	\label{eq5}
 	\begin{aligned}
 		\pi \left( x\left( t \right) \right)=-{{K}_{i,\text{CBF}}}x\left( t \right)+{{F}_{i}}y_{\lim }^{\min /\max },
 	\end{aligned}
 \end{equation}
 with the CBF feedback gain matrix  ${{K}_{i,\text{CBF}}}$ and the CBF command matrix ${{F}_{i}}$ switching to accommodate constraints, such that the overall control input $u\left( t \right)$ is continuous, closed-loop stability is guaranteed, and the limited output constraints~\eqref{eq2} are satisfied forward in time. The two key steps in deriving the CBF augmentation~\eqref{eq5} are the derivation of modified constraints and the derivation of an analytical solution to satisfy the modified constraints, which are discussed next.
 
The key idea behind the two steps is: (i) Derive modified constraints where the control input appears explicitly such that a dynamical system satisfying the modified constraints also satisfies the original operational constraints~\eqref{eq2}, and (ii) compute the min-norm control input using the modified constraints that renders the dynamical system forward invariant with respect to the modified constraints, and consequently also with respect to the original operational constraints~\eqref{eq2}.

 \subsection{Derivation of Modified Constraints}
 \label{ssec3a}
 
 Let the relative degree of the $i$-th limited output  ${{y}_{\lim ,i}}\left( t \right)$ be ${{r}_{i}}\ge 1$ for $i=1,\,\ldots ,\,m$, where relative degree refers to the number of time ${{y}_{\lim ,i}}\left( t \right)$ needs to be differentiated for the control augmentation signal to appear explicitly. 
 Suppose that ${{y}_{\lim }}$ has a vector relative degree $r=(\begin{matrix}
 	{{r}_{1}} & \ldots  & {{r}_{m}}  \\
 \end{matrix})$, i.e.,
 \begin{equation}
 \label{eq6}
 \begin{aligned}
 	\left[ \left\| {{\nabla }_{\pi }}\left( y_{\lim ,i}^{\left( k \right)} \right) \right\|=0,\ \forall 1\le k\le {{r}_{i}}-1 \right]\wedge \left[ \left\| {{\nabla }_{\pi }}\left( y_{\lim ,i}^{\left( {{r}_{i}} \right)} \right) \right\|\ne 0 \right].
 \end{aligned}
 \end{equation}
 Then, for a dynamical system with operational constraints to be CBF-able, the control-to-limited output sensitivity matrix ${{H}_{\pi }}\in {{R}^{m\times m}}$ must be nonsingular, see Assumption~\ref{assum2},
 \begin{equation}
 	\label{eq7}
 	\begin{aligned}
 		{{H}_{\pi }}=\left[ \begin{matrix}
 			{{\left( {{C}_{\lim }} \right)}_{1}}A_{{}}^{{{r}_{1}}-1}B  \\
 			\vdots   \\
 			{{\left( {{C}_{\lim }} \right)}_{m}}A_{{}}^{{{r}_{m}}-1}B  \\
 		\end{matrix} \right],\quad \det \left( {{H}_{\pi }} \right)\ne 0.
 	\end{aligned}
 \end{equation}
 For every limited output $i=1,...,m$, the vector relative degree definition is used to differentiate the operational constraints ${{r}_{i}}$ times as follows. Consider a stable polynomial of order ${{r}_{i}}$, with the real roots ${{\left\{ {{\lambda }_{i\,j}} \right\}}_{j=1,\,\ldots ,\,{{r}_{i}}}}$ with ${{\lambda }_{ij}}<0$
 \begin{equation}
 	\label{eq8}
 	\begin{aligned}
 		{{\phi }_{i}}\left( s \right)=\prod\limits_{j=1}^{{{r}_{i}}}{\left( s-{{\lambda }_{i\,j}} \right)}=\sum\limits_{j=0}^{{{r}_{i}}}{{{c}_{i\,j}}\,{{s}^{j}}},
 	\end{aligned}
 \end{equation}
 where ${{c}_{i\,j}}$ denotes the ${{j}^{\text{th}}}$ coefficient of the ${{i}^{\text{th}}}$ polynomial, and by definition, ${{c}_{i\,0}}>0$, for every $i=1,\,\ldots ,\,m$. The stable polynomials in~\eqref{eq8} are used to modify the constraints as
 \begin{equation}
 	\label{eq9}
 	\begin{matrix}
 		{{Y}_{\lim }}\left( t \right)
 		=\Phi \left( s \right){{y}_{\lim }}\left( t \right) ,\quad
 		\Phi \left( s \right)={\rm diag}\left(\begin{bmatrix}
 			{{\phi }_{1}}\left( s \right) & \ldots  &  {{\phi }_{m}}\left( s \right)
 		\end{bmatrix}\right)
 		\\ 
 		H\left( x( t),{{u}_{\rm bl}}\left( t \right),\pi \left( t \right) \right)=\left[ \begin{matrix}
 			{{H}_{\min }}\left( x\left( t \right),{{u}_{\text{bl}}}\left( t \right),\pi \left( t \right) \right)  \\
 			{{H}_{\max }}\left( x\left( t \right),{{u}_{\text{bl}}}\left( t \right),\pi \left( t \right) \right)  \\
 		\end{matrix} \right]
 		=\left[ \begin{matrix}
 			\Phi \left( s \right){{h}_{\min }}\left( x\left( t \right) \right)  \\
 			\Phi \left( s \right){{h}_{\max }}\left( x\left( t \right) \right)  \\
 		\end{matrix} \right]= \begin{bmatrix}
 			\Phi \left( s \right)
 			\left( y_{\lim }^{\min }-y_{\lim }\left( t \right) \right)  
 			\\
 			\Phi \left( s \right)
 			\left( y_{\lim }\left( t \right)-y_{\lim }^{\max } \right) 
 		\end{bmatrix} \le 0
 	\end{matrix}
 \end{equation}
 where the stable polynomials ${{\phi }_{i}}\left( s \right)$ are treated as differentiation operators with respect to $s$. Then,  Lemma~\ref{lem1acc} with proof in the appendix shows that the modified output ${{Y}_{\lim }}\left( t \right)$ and modified constraints can be written as
 \begin{equation}
 	\label{eq10}
 	\begin{aligned}
 		\begin{matrix}
 			{{Y}_{\lim }}\left( t \right)={{H}_{x}}x\left( t \right)+{{H}_{\pi }}\left( {{u}_{\text{bl}}}\left( t \right)+\pi \left( t \right) \right) \\ 
 			\left[ \begin{matrix}
 				{{H}_{\min }}\left( x\left( t \right),{{u}_{\text{bl}}}\left( t \right),\pi \left( t \right) \right)  \\
 				{{H}_{\max }}\left( x\left( t \right),{{u}_{\text{bl}}}\left( t \right),\pi \left( t \right) \right)  \\
 			\end{matrix} \right]=\left[ \begin{matrix}
 				-{{Y}_{\lim }}\left( t \right)+{{\alpha }_{\pi }}\,y_{\lim }^{\min }  \\
 				\,\,\,{{Y}_{\lim }}\left( t \right)-{{\alpha }_{\pi }}\,y_{\lim }^{\max }  \\
 			\end{matrix} \right]=\left[ \begin{matrix}
 				-{{H}_{\pi }}  \\
 				{{H}_{\pi }}  \\
 			\end{matrix} \right]\pi \left( t \right)+\left[ \begin{matrix}
 				\Delta {{H}_{\min }}\left( x\left( t \right),{{u}_{\text{bl}}}\left( t \right) \right)  \\
 				\Delta {{H}_{\max }}\left( x\left( t \right),{{u}_{\text{bl}}}\left( t \right) \right)  \\
 			\end{matrix} \right]\le 0 \\ 
 		\end{matrix}
 	\end{aligned}
 \end{equation}
 with
 \begin{equation}
 	\label{eq11}
 	\begin{aligned} 
 		\begin{matrix}
 			\Delta {{H}_{\min }}\left( x\left( t \right),{{u}_{\text{bl}}}\left( t \right) \right)=-{{H}_{x}}x\left( t \right)-{{H}_{\pi }}{{u}_{\text{bl}}}\left( t \right)+{{\alpha }_{\pi }}\,y_{\lim }^{\min } \\ 
 			\Delta {{H}_{\max }}\left( x\left( t \right),{{u}_{\text{bl}}}\left( t \right) \right)=\,{{H}_{x}}x\left( t \right)+{{H}_{\pi }}{{u}_{\text{bl}}}\left( t \right)-{{\alpha }_{\pi }}\,y_{\lim }^{\max } \\ 
 		\end{matrix}
 	\end{aligned}
 \end{equation}
 and ${{H}_{x}}$ and ${{H}_{\pi }}$ in~\eqref{eq4} are the system state and control sensitivity matrices respectively, and the diagonal matrix ${{\alpha }_{\pi }}$ with positive elements ${{\alpha }_{\pi }}=\text{diag}( [ \begin{matrix}
 	{{c}_{1\,0}} & \ldots  & {{c}_{m\,0}}  \\
 \end{matrix}] )\in {{R}^{m\times m}}$ with ${{c}_{i\,0}}=\prod\nolimits_{j=1}^{r}{\left( -{{\lambda }_{i\,j}} \right)>0}$ for all $i=1,\,\ldots ,\,m.$

 \begin{lem}
 	\label{lem1acc}
 	Suppose that ${{y}_{\lim }}$ has vector relative degree $r=(\begin{matrix}
 		{{r}_{1}} & \ldots  & {{r}_{m}}  \\
 	\end{matrix})$, with ${{r}_{i}}\ge 1$ for all $i.$ Then, using the stable polynomials \eqref{eq8} and \eqref{eq9} as differentiation operators applied to the operational constraints~\eqref{eq2}, results in the modified operational constraints \eqref{eq10} with ${{H}_{\pi }}$ and ${{H}_{x}}$ in \eqref{eq4}.
 \end{lem}

 \subsection{Derivation of Analytical Solution from Quadratic Program}
 
 Since the control augmentation appears explicitly in the modified constraints~\eqref{eq10}, it can directly be used for control design. 
 Motivated by the min-norm controller design method \cite{FreemanKokotovic1996}, consider the following QP \cite{BoydVandenberghe2004}:
 \begin{equation}
 	\label{eq12}
 	\begin{aligned}  
 		\begin{matrix}
 			\text{Cost}:J\left( v,w \right)=\left( {{\pi }^{T}}{{R}_{\pi }}\pi  \right)\to \underset{\pi }{\mathop{\min }}\, \\ 
 			\text{Constraints}:H\left( x,\,{u_{\rm bl}},\,\pi  \right)=\left[ \begin{matrix}
 				{{H}_{\min }}\left( x,{u_{\text{bl}}},\pi  \right)  \\
 				{{H}_{\max }}\left( x,{u_{\text{bl}}},\pi  \right)  \\
 			\end{matrix} \right]=\left[ \begin{matrix}
 				-{{H}_{\pi }}  \\
 				{{H}_{\pi }}  \\
 			\end{matrix} \right]\pi +\left[ \begin{matrix}
 				\Delta {{H}_{\min }}\left( x,{u_{\text{bl}}} \right)  \\
 				\Delta {{H}_{\max }}\left( x,{u_{\text{bl}}} \right)  \\
 			\end{matrix} \right]\le 0 \\ 
 		\end{matrix}
 	\end{aligned}
 \end{equation}
 i.e., we seek the min-norm control augmentation that only becomes active (meaning $\pi \ne 0$) if needed to satisfy the modified constraints in~\eqref{eq12}. Hence, the modified constraints are constructed such than any $\pi $ satisfying~\eqref{eq12} also makes the dynamical system forward invariant with respect to the operational constraints. Further, choosing the cost function weight ${{R}_{\pi }}=H_{\pi }^{T}{{H}_{\pi }}$,~\eqref{eq12} has an explicit closed-form solution, which is formally shown in Theorem~\ref{thm1}. The state-dependent control augmentation policy optimizing the QP in~\eqref{eq12} is given by
 \begin{equation}
 	\label{eq13}
 	\begin{aligned}  
 			\pi \left( x\left( t \right) \right)& = H_{\pi }^{-1}\left( \max \left( {{0}_{m\times 1}},\ \Delta {{H}_{\min }}\left( x\left( t \right),\,\,{u_{\text{bl}}}\left( t \right) \right) \right)-\max \left( {{0}_{m\times 1}},\ \Delta {{H}_{\max }}\left( x\left( t \right),\,\,{u_{\text{bl}}}\left( t \right) \right) \right) \right) \\ 
 			& =H_{\pi }^{-1}\left\{ \begin{matrix}
 				& \left( -{{H}_{x}}x\left( t \right)-{{H}_{\pi }}{u_{\text{bl}}}\left( t \right)+{{\alpha }_{\pi }}y_{\lim }^{\min } \right),
 				& \text{if}\ \left( -{{H}_{x}}x\left( t \right)-{{H}_{\pi }}{u_{\text{bl}}}\left( t \right)+{{\alpha }_{\pi }}y_{\lim }^{\min } \right)>0 \\ 
 				& -\left( {{H}_{x}}x\left( t \right)+{{H}_{\pi }}{u_{\text{bl}}}\left( t \right)-{{\alpha }_{\pi }}y_{\lim }^{\max } \right),
 				& \text{if}\ \left( {{H}_{x}}x\left( t \right)+{{H}_{\pi }}{u_{\text{bl}}}\left( t \right)-{{\alpha }_{\pi }}y_{\lim }^{\max } \right)>0 \\ 
 				& 0, 
 				& \text{otherwise} \\ 
 			\end{matrix} \right.  
 	\end{aligned}
 \end{equation}
 with ${{H}_{x}}$ and ${{H}_{\pi }}$ as in~\eqref{eq4}. Theorem~\ref{thm1} also shows forward invariance with respect to both the modified constraints~\eqref{eq10} and consequently the operational constraints. 
 
 \begin{thm}
 	\label{thm1}
 	Consider the controllable LTI system~\eqref{eq1} and the QP~\eqref{eq12} with the positive definite cost ${{R}_{\pi }}=H_{\pi }^{T}{{H}_{\pi }}$.  Then the CBF augmentation $\pi $ in~\eqref{eq13} is optimal with respect to the QP in~\eqref{eq12}.
 \end{thm}
 
 Eq.~\eqref{eq13} can also be expressed in the form~\eqref{eq5}. To see this, we reformulate the control augmentation solution. Let 
 \begin{equation}
 	\label{eq14}
 	\begin{aligned}   
 		\delta \left( x\left( t \right) \right)=\text{diag}\left( [
 			{{\delta }_{1}}\left( x\left( t \right) \right) \quad \ldots  \quad {{\delta }_{m}}\left( x\left( t \right) \right) 
 		] \right)\in {{R}^{m\times m}}
 	\end{aligned}
 \end{equation}
 defining a diagonal state-dependent matrix, whose binary-valued elements are given by the conditions in~\eqref{eq13}
 \begin{equation}
 	\label{eq15}
 	\begin{aligned}   
 		[{{\delta }_{i}}\left( x\left( t \right) \right)=\left\{ \begin{matrix}
 			& 1,& \text{if}\ \left[ {{\left( -{{H}_{x}}x\left( t \right)-{{H}_{\pi }}{u_{\text{bl}}}\left( t \right)+{{\alpha }_{\pi }}\,y_{\lim }^{\min } \right)}_{i}}>0 \right]\vee \left[ {{\left( {{H}_{x}}x\left( t \right)+{{H}_{\pi }}{u_{\text{bl}}}\left( t \right)-{{\alpha }_{\pi }}\,y_{\lim }^{\max } \right)}_{i}}>0 \right] \\ 
 			& 0,& \text{otherwise} \\ 
 		\end{matrix} \right.
 	\end{aligned}
 \end{equation}
 $\forall i=1,\,\ldots ,\,m$. Hence, the notation ${{\delta }_{i}}\left( x\left( t \right) \right)$ represents a continuous switching logic based on active constraints, and, by definition, $\left\| \delta \left( x\left( t \right) \right) \right\|\le 1$ uniformly in $x\left( t \right)$. Then, the total controller shows a modified baseline control part, a CBF feedback part, and a CBF command part:
 \begin{equation}
 	\label{eq16}
 	\begin{aligned}   
 			{u_{\text{bl}}}\left( t \right)+\pi \left( t \right)
 			& ={u_{\text{bl}}}\left( t \right)-H_{\pi }^{-1}\delta \left( x\left( t \right) \right)\left( {{H}_{x}}x\left( t \right)+{{H}_{\pi }}\,{u_{\text{bl}}}\left( t \right)-{{\alpha }_{\pi }}y_{\lim }^{\min /\max } \right) \\ 
 			& =\underbrace{\left( {{I}_{\left( m \right)}}-H_{\pi }^{-1}\delta \left( x\left( t \right) \right){{H}_{\pi}} \right){{u}_{\text{bl}}}\left( t \right)}_{\text{Modified Baseline Control}}-\underbrace{H_{\pi }^{-1}\delta \left( x\left( t \right) \right){{H}_{x}}x\left( t \right)}_{\text{CBF}\,\text{Feedback}}+\underbrace{H_{\pi }^{-1}\delta \left( x\left( t \right) \right){{\alpha }_{\pi }}y_{\lim }^{\min /\max }}_{\text{CBF}\,\text{Command}}  
 	\end{aligned}
 \end{equation}
 where the ${{i}^{\text{th}}}$ component of the constant CBF command vector $y_{\lim }^{\min /\max }\in {{R}^{m}}$ is defined as 
 \begin{equation}
 	\label{eq17}
 	\begin{aligned}   
 		{{\left( y_{\lim }^{\min /\max } \right)}_{i}}=\left\{ \begin{matrix}
 			& {{\left( y_{\lim }^{\min } \right)}_{i}} & \text{if}\ {{\left( -{{H}_{x}}x\left( t \right)-{{H}_{\pi }}{u_{\text{bl}}}\left( t \right)+{{\alpha }_{\pi }}\,y_{\lim }^{\min } \right)}_{i}}>0 \\ 
 			& {{\left( y_{\lim }^{\max } \right)}_{i}} & \text{if}\ {{\left( {{H}_{x}}x\left( t \right)+{{H}_{\pi }}{u_{\text{bl}}}\left( t \right)-{{\alpha }_{\pi }}\,y_{\lim }^{\max } \right)}_{i}}>0 \\ 
 			& 0 & \text{otherwise} \\ 
 		\end{matrix} \right.
 	\end{aligned}
 \end{equation} 
 It can thus be seen that~\eqref{eq5} can be written as
 \begin{equation}
 	\label{eq18}
 	\begin{aligned}   
 		\pi \left( t \right)=-\underbrace{H_{\pi }^{-1}\delta \left( x\left( t \right) \right)\left( {{H}_{x}}-{{H}_{\pi}}K \right)}_{{{K}_{i,\text{CBF}}}}x\left( t \right)+\underbrace{H_{\pi }^{-1}\delta \left( x\left( t \right) \right){{\alpha }_{\pi }}}_{{{F}_{i}}}y_{\lim }^{\min /\max }=-{{K}_{i,\text{CBF}}}x\left( t \right)+{{F}_{i}}y_{\lim }^{\min /\max }.
 	\end{aligned}
 \end{equation}  
 The equivalent form in~\eqref{eq18} highlights that the proposed control augmentation remains closed-loop even when constraints are active. Finally, it can be seen that “infeasibilities/constraint violations” occurring, e.g., due to noise or disturbances, are systematically handled. While optimization-based methods need to pay special attention to address infeasibilities, the proposed formulation uses a QP only for design and thus,~\eqref{eq18} provides a valid solution at all times.

 \subsection{Constraint Satisfaction and Forward Invariance}
 \label{ssec:constraint}
 
 Lemma~\ref{lem2} extends the Nagumo Theorem \cite{Nagumo1942,MennerLavretsky2024} showing that if a dynamical system is invariant forward in time with respect to the modified constraints~\eqref{eq10}, then it is also invariant forward in time with respect to the original constraints~\eqref{eq2}. Thus, designing the control augmentation to satisfy~\eqref{eq10} will enforce~\eqref{eq2}, which is shown in Theorem~\ref{thm2}. 
 
 \begin{lem}
 	\label{lem2}
 	Consider the system of differential equations,
 			$\dot{x}=f( x ),\, x\in {{R}^{n}}$
 	with $f:{{R}^{n}}\to {{R}^{n}}$ that is Lipchitz in $x$. Let $h:{{R}^{n}}\to R$ be an $r-$times continuously differentiable function. Consider the non-empty subset in ${{R}^{n}},$ $H=\left\{ x\in {{R}^{n}}:h\left( x \right)\le 0 \right\}.$ Suppose that for all initial conditions $x\left( 0 \right)={{x}_{0}}\in H$ the system trajectories are well-defined forward in time. Let ${{\phi }_{r}}\left( s \right)=\prod\nolimits_{j=1}^{r}{\left( s-{{\lambda }_{j}} \right)}=\sum\nolimits_{j=0}^{r}{{{c}_{j}}\,{{s}^{j}}}$ be a stable polynomial of order $r\ge 1$, with real negative ${{\lambda }_{j}}$. 
 	Consider the set ${{\tilde{H}}_{r}}=\left\{ x\in {{R}^{n}}:{{{\tilde{h}}}_{r}}\left( x \right)\le 0 \right\}$ with ${{\tilde{h}}_{r}}\left( x \right)={{\phi }_{r}}\left( s \right)h\left( x \right)=\sum\nolimits_{j=0}^{r}{{{c}_{j}}\,{{h}^{\left( j \right)}}\left( x\left( t \right) \right)},$	and ${{h}^{\left( j \right)}}(x)$ is the ${{j}^{\text{th}}}$ time derivative of $h(x)$ along the system trajectories. If ${{\tilde{H}}_{r}}$ is forward invariant, then so is $H,$
 	\begin{equation}
 		\label{eq23}
 		\begin{aligned} 
 			\left[ x\left( t \right)\in {{{\tilde{H}}}_{r}} \right]\Rightarrow \left[ x\left( t \right)\in H \right],\ \forall t\ge 0.
 		\end{aligned}
 	\end{equation}
 \end{lem}
 
\begin{thm}
	\label{thm2}
	 Consider the controllable LTI system~\eqref{eq1}. Suppose that the limited output ${{y}_{\lim }}$ has vector relative degree $r=(\begin{matrix}
		{{r}_{1}} & \ldots  & {{r}_{m}}  \\
	\end{matrix})$. Then, for $x({{t}_{0}}=0)\in {{X}_{0}}$ with the set ${{X}_{0}}=\left\{ x:H(x,\pi )\le 0 \right\}$ defined by~\eqref{eq10}, the CBF augmentation~\eqref{eq13} enforces the operational constraints~\eqref{eq2} imposed on the limited output ${{y}_{\lim }}$ forward in time, $t\ge {{t}_{0}}.$
\end{thm}

 \subsection{CBF-ability and Closed-Loop Stability}
 
 Suppose that the eigenvalues ${{\lambda }_{i\,j}}$ of the stable polynomial in~\eqref{eq8} are real negative and chosen such that ${{A}_{{\rm CBF}}}$ is Hurwitz for any diagonal matrix $\delta (t)$ with binary diagonal elements, 0 or 1. Then, the closed-loop system with the baseline controller ${u_{\rm bl}}$ and the optimal CBF augmentation $\pi $ remains stable in the sense of Lyapunov, throughout the operational domain, including the set boundary.
 Next, we derive the formulation used in Definition~\ref{def1} (CBF-ability). Since the control augmentation solution provides a continuous, piecewise-linear solution, closed-loop stability and boundedness can be assessed. Consider the modified constrained output 
 ${{Y}_{\lim }}\left( t \right)={{H}_{x}}x\left( t \right)+{{H}_{\pi }}u\left( t \right)$ in~\eqref{eq10}. By construction, the control augmentation enforces
 \begin{equation}
 	\label{eq19}
 	\begin{aligned}  {{\alpha }_{\pi }}\,y_{\lim }^{\min }\le {{Y}_{\lim }}\left( t \right)\le {{\alpha }_{\pi }}\,y_{\lim }^{\max } .	\end{aligned}
 \end{equation}
 We can express the total control input as
 $u\left( t \right)=H_{\pi }^{-1}\left( {{Y}_{\lim }}\left( t \right)-{{H}_{x}}x\left( t \right) \right)$, with the system dynamics in~\eqref{eq1} given by 
 \begin{equation}
 	\label{eq21}
 	\begin{aligned} 
 		\dot{x}\left( t \right)&=Ax\left( t \right)+Bu\left( t \right)=Ax\left( t \right)+BH_{\pi }^{-1}\left( {{Y}_{\lim }}\left( t \right)-{{H}_{x}}x\left( t \right) \right) 
 		=\left( A-BH_{\pi }^{-1}{{H}_{x}} \right)x\left( t \right)+BH_{\pi }^{-1}{{Y}_{\lim }}\left( t \right)  .
 	\end{aligned} 
 \end{equation}
 Hence, the closed-loop system dynamics for baseline control and control augmentation are driven by the bounded signal ${{Y}_{\lim }}\left( t \right)$. Since the eigenvalues of $A-BH_{\pi }^{-1}{{H}_{x}}$ do not depend on the baseline control design, the ability to apply the proposed augmentation technique can directly be checked. Further, the tuning parameters of the control augmentation (the eigenvalues in~\eqref{eq8}) need to be chosen such that ${{A}_{\text{CBF}}}=A-BH_{\pi }^{-1}{{H}_{x}}$ is Hurwitz.
 Lemma~\ref{lem1} shows necessary and sufficient condition on the plant parameters with relative degree one limited outputs (independent of the baseline control) for which a CBF augmentation can be applied such that closed-loop stability is maintained.

 \begin{lem}
 	\label{lem1}
 	Consider the LTI system~\eqref{eq1} with all limited outputs having relative degree one, ${{r}_{i}}=1$. Let ${{B}^{\bot }}$ and ${{C}^{\bot }}$ define null spaces of $B$ and $C$ such that ${{B}^{\bot }}B={{0}_{m\times m}},C{{C}^{\bot }}={{0}_{m\times m}},$ and ${{B}^{\bot }}{{C}^{\bot }}={{I}_{m}}$. The following statements hold:
 	\begin{enumerate}
 		\item A necessary and sufficient condition for CBF-ability on the plant parameters is ${{B}^{\bot }}A{{C}^{\bot }}$ being Hurwitz.
 		\item If there exists a set of CBF tuning parameters ${{\alpha }_{\pi ,ij}}=-{{\lambda }_{ij}}>0$ with ${{\lambda }_{ij}}$ as in~\eqref{eq8} for which ${{A}_{\text{CBF}}}=A-BH_{\pi }^{-1}{{H}_{x}}$ is Hurwitz, then ${{A}_{\text{CBF}}}$ is Hurwitz for any ${{\alpha }_{\pi ,ij}}=-{{\lambda }_{ij}}>0$. 
 		\item If there exists a set of ${{\alpha }_{\pi ,ij}}=-{{\lambda }_{ij}}>0$ with ${{\lambda }_{ij}}$ as in~\eqref{eq8} for which ${{A}_{\text{CBF}}}$ is not Hurwitz, then ${{A}_{\text{CBF}}}$ is not Hurwitz for any ${{\alpha }_{\pi ,ij}}=-{{\lambda }_{ij}}>0$. 
 	\end{enumerate}
 \end{lem}

 \subsubsection{CBF-ability toy example}
 
\begin{figure}[t]  
 	\centerline{ 
 		\includegraphics[width=7.5cm,trim={0 5.0cm 0 0},clip]{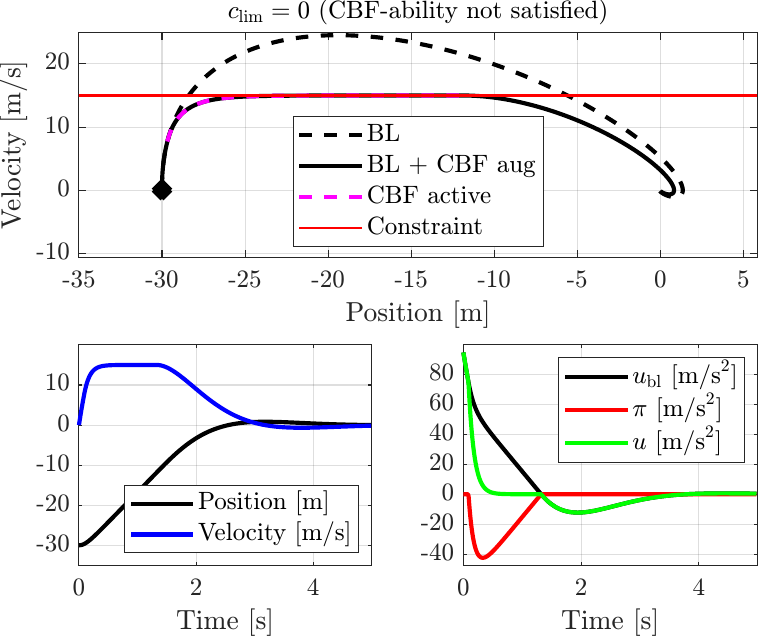}\hspace*{0.5cm}
 		\includegraphics[width=7.5cm,trim={0 5.0cm 0 0},clip]{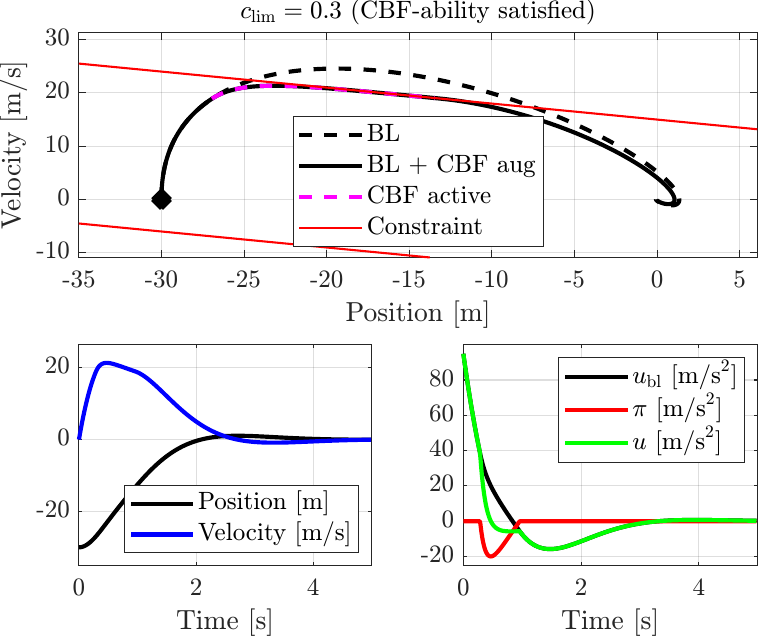} 
 	} 
 	\centerline{ 
 		\includegraphics[width=7.5cm,trim={0 5.3cm 0 0},clip]{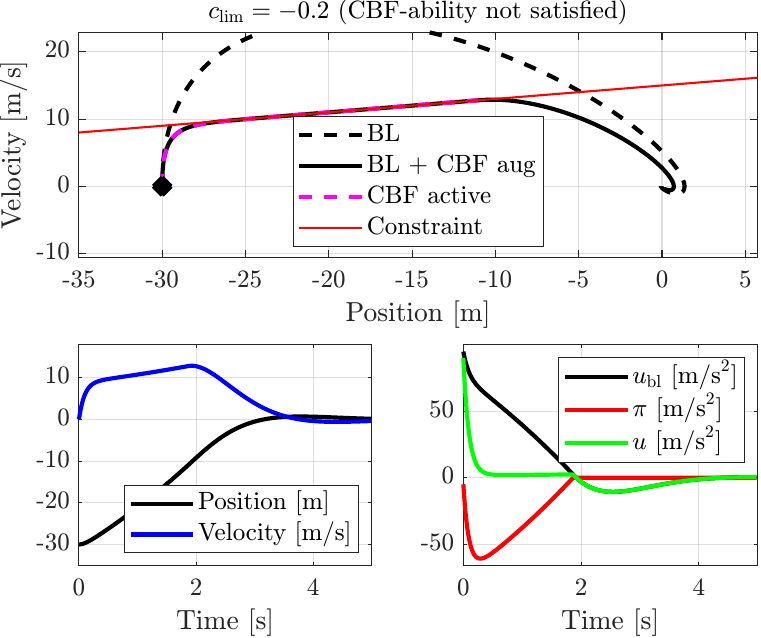}\hspace*{0.5cm}
 		\includegraphics[width=7.5cm,trim={0 5.3cm 0 0},clip]{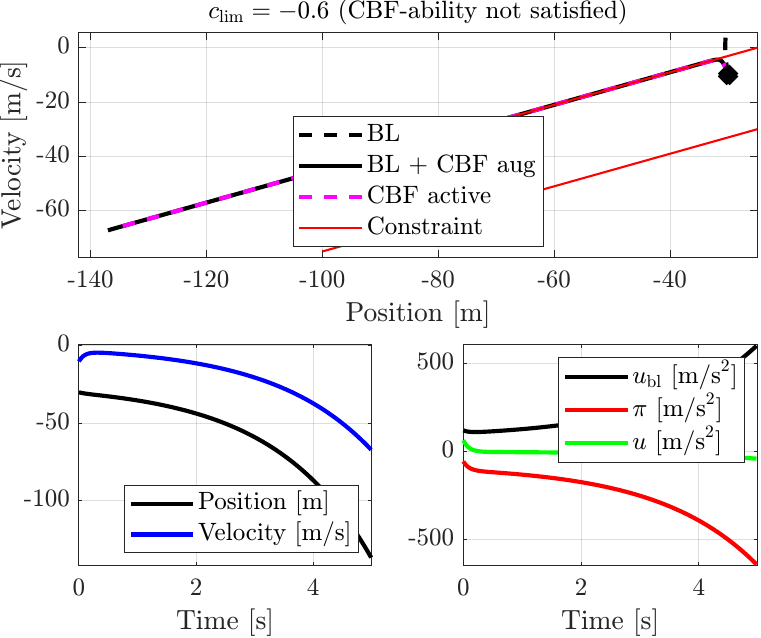}
 	} \caption{Closed-loop trajectories for double integrator with differing limited output. 
 	} 
 	\label{fig1} 
 \end{figure}

 Consider a double integrator model with a baseline LQR controller with position and velocity states, $p\left( t \right)$ and $v\left( t \right)$. The CBF augmentation is designed to keep the limited output bounded with ${{y}_{\max }}=15,{{y}_{\min }}=-15,$
 \begin{equation}
 	\label{eq53}
 	\begin{aligned} 
 		\begin{matrix}
 			\begin{bmatrix}
 				\dot{p}\left( t \right) \\ 
 				\dot{v}\left( t \right) \\ 
 			\end{bmatrix} = \begin{bmatrix}
 				0 & 1  \\
 				0 & 0  \\
 			\end{bmatrix} \begin{bmatrix}
 				p\left( t \right) \\ 
 				v\left( t \right) \\ 
 			\end{bmatrix}+ \begin{bmatrix}
 				0 \\ 
 				1 \\ 
 			\end{bmatrix} u\left( t \right) ,\quad 
 			{{y}_{\lim }}= \begin{bmatrix}
 				{{c}_{\lim ,1}} & 1  \\
 			\end{bmatrix} \begin{bmatrix}
 				p\left( t \right) \\ 
 				v\left( t \right) \\ 
 			\end{bmatrix} 
 		\end{matrix}
 	\end{aligned}
 \end{equation}
 with the control input $u\left( t \right)={u_{\text{bl}}}\left( t \right)+\pi \left( t \right)=\left[ 3.16 \quad 2.51 \right] [
 	p\left( t \right) \quad
 	v\left( t \right) 
 ]^T+\pi \left( t \right)$. 
 In the following ${{c}_{\lim ,1}}$ is varied to highlight CBF-ability properties. As shown in Lemma~\ref{lem1}, the null spaces for~\eqref{eq53} needed to show CBF-ability are given by ${{B}^{\bot }}=\left[ 1 \quad 0 \right],{{C}^{\bot }}= \left[ 1 \quad -{{c}_{\lim ,1}} \right]^{T}$. Thus, the necessary and sufficient condition for CBF-ability in Lemma~\ref{lem1} is 
${{B}^{\bot }}A{{C}^{\bot }}=-{{c}_{\lim ,1}}<0$. 
 Hence, any positive ${{c}_{\lim ,1}}$ will guarantee closed-loop stability/boundedness.  Fig.~\ref{fig1} highlights CBF-ability for different choices of ${{c}_{\lim ,1}}$. It shows that the closed-loop system can only be guaranteed to be bounded if the CBF-ability condition is satisfied. If CBF-ability is not satisfied, the system's states may become unbounded.
 In all scenarios, the closed-loop trajectories approach and ride the constraint. 
 But in the bottom right plot, the closed-loop trajectory can never leave the constraint upon entering, which is due to position and velocity decreasing simultaneously. This behavior is formalized in this paper with the notion of CBF-ability.
 
 \subsection{Robustness Margins and Relative Stability}
 
 For any servo-control design framework, robustness margins need to be analyzed. Here, we show how to assess gain and phase margins at the control input breakpoint, where the baseline control ${u_{\text{bl}}}\left( t \right)$ and the augmentation signal $\pi \left( t \right)$ are added to form the total control input $u\left( t \right)$. 
Other breakpoints follow analogously.
Using the notation in~\eqref{eq16}, the total control input can be written in terms of the continuous switching logic~\eqref{eq5}	and the closed-loop system yields
 \begin{equation}
 	\label{eq24}
 	\begin{aligned} 
 		\dot{x}\left( t \right)=Ax+Bu=Ax-B\left( \left( K+{{K}_{i,\text{CBF}}} \right)x\left( t \right)-{{F}_{i}}y_{\lim }^{\min /\max } \right).
 		\end{aligned}
 	\end{equation}
 The system loop gain transfer function matrix ${{L}_u}\left( s \right)$ can be computed with the total servo-controller~\eqref{eq16}, while zeroing out command/constant terms $y_{\lim }^{\min /\max }$, ${u_{\text{out}}}\left( t \right)=-\left( K+{{K}_{i,\text{CBF}}} \right)x\left( t \right)$. 
Using $x\left( t \right)={{\left( s\,{{I}_{n}}-A \right)}^{-1}}B{u_{\text{in}}}\left( t \right)$, we can state the loop gain $-{{L}_u}\left( s;\,\delta  \right)$ to compute SISO and MIMO gain and phase margins \cite{Lavretsky2024} for all possible combinations of the binary-valued diagonal elements of $\delta$, where $\delta $ is assumed to be constant diagonal matrix for this analysis,
 \begin{equation}
 	\label{eq25}
 	\begin{aligned} 
 		{u_{\text{out}}}\left( t \right)=-\left( K+{{K}_{i,\text{CBF}}} \right){{\left( s\,{{I}_{n}}-A \right)}^{-1}}B{u_{\text{in}}}\left( t \right)=-{{L}_u}\left( s;\,\delta  \right){u_{\text{in}}}\left( t \right).
 		\end{aligned}
 	\end{equation}

  \section{Augmentation Design for Proportional-Integral Controllers}
  Consider the stabilizable MIMO LTI dynamical system,
  \begin{equation}
  	\label{eq37}
  	\begin{aligned} 
  		\begin{matrix}
  	{{{\dot{x}}}_{p}}\left( t \right)={{A}_{p}}\,{{x}_{p}}\left( t \right)+{{B}_{p}}u\left( t \right) \\ 
  	y\left( t \right)={{C}_{p}}{{x}_{p}}\left( t \right)+{{D}_{p}}\,u\left( t \right) \\ 
  	{{y}_{\text{reg}}}\left( t \right)={{C}_{p\,\,\text{reg}}}\,{{x}_{p}}\left( t \right)+{{D}_{p\,\,\text{reg}}}\,u\left( t \right) \\ 
  	{{z}_{\lim }}\left( t \right)={{C}_{p\,\lim }}\,{{x}_{p}}\left( t \right) \\ 
  \end{matrix}
\end{aligned}
\end{equation}
  where ${{x}_{p}}\left( t \right)\in {{R}^{{{n}_{p}}}}$ is the state vector, $y\left( t \right)\in {{R}^{{{n}_{y}}}}$ is the vector of measured outputs,${{y}_{\text{reg}}}\left( t \right)\in {{R}^{m}}$ is the vector of regulated outputs, and ${{z}_{\text{lim}}}\left( t \right)\in {{R}^{m}}$ is the vector of limited outputs. Here, we apply the generic design in Section III to a PI servo-controller with robust tracking of external commands ${{y}_{\text{cmd}}}$, while min/max operational constraints on the system control input $u\left( t \right)$ and on the selected limited outputs ${{z}_{\lim }}\left( t \right)$ are satisfied for all times, component-wise,
  \begin{equation}
  	\label{eq38}
  	\begin{aligned} 
  		\begin{matrix}
  	{u^{\min }}\le u\left( t \right)\le {u^{\max }} \\ 
  	z_{\lim }^{\min }\le {{z}_{\lim }}\left( t \right)\le z_{\lim }^{\max }. \\ 
  \end{matrix}
\end{aligned}
\end{equation}

  Consider the control command,
  \begin{equation}
  	\label{eq39}
  	\begin{aligned} 
  		u\left( t \right)={u_{\text{bl}}}\left( t \right)+w\left( t \right)
  \end{aligned}
\end{equation}
  with the proportional control augmentation signal $w\left( t \right)$. For robust tracking of external commands ${{y}_{\text{cmd}}}$ and operating in the presence of operational constraints, we use an integrated output tracking error ${{e}_{yI}}\left( t \right)$ augmented with the anti-windup (AW) control modification term $v\left( t \right)\in {{R}^{m}}$,
  \begin{equation}
  	\label{eq40}
  	\begin{aligned} 
  	& \dot{e}_{y,I}\left( t \right)={e}_{y}\left( t \right)+v\left( t \right) \\ 
  	& {{e}_{y}}\left( t \right)={{y}_{\text{reg}}}\left( t \right)-{{y}_{\text{cmd}}}  
\end{aligned}
\end{equation}
with the output tracking error signal ${{e}_{y}}\left( t \right)$. 
Hence for the PI case, we design a feedback control augmentation policy $w\left( t \right)$ and $v\left( t \right)$ such that the system’s regulated output ${{y}_{\text{reg}}}\left( t \right)$ tracks external commands ${{y}_{\text{cmd}}}$ to the extent possible, while the control input $u\left( t \right)$ and the limited output ${{z}_{\lim }}\left( t \right)$ evolve within their operational constraints~\eqref{eq38}. Further, the state ${{x}_{p}}\left( t \right)$ and integrated output error ${{e}_{y\,I}}\left( t \right)$ remain bounded for all times.
For the PI controller design, the system is reformulated such that the generic control augmentation in Section~\ref{sec3} can be applied directly. First, augmenting the system dynamics in~\eqref{eq37} with the error dynamics in~\eqref{eq40} gives the $n=({{n}_{p}}+m)-$dimensional extended system
  \begin{equation}
  	\label{eq41}
  	\begin{aligned} 
  		\begin{matrix}
  	\underbrace{\left[ \begin{matrix}
  			{{{\dot{e}}}_{y\,I}}\left( t \right)  \\
  			{{{\dot{x}}}_{p}}\left( t \right)  \\
  		\end{matrix} \right]}_{\dot{x}\left( t \right)}=\underbrace{\left[ \begin{matrix}
  			{{0}_{m\times m}} & {{C}_{p\,\,\text{reg}}}\,  \\
  			{{0}_{n\times m}} & {{A}_{p}}  \\
  		\end{matrix} \right]}_{A}\underbrace{\left[ \begin{matrix}
  			{{e}_{y\,I}}\left( t \right)  \\
  			{{x}_{p}}\left( t \right)  \\
  		\end{matrix} \right]}_{x\left( t \right)}+\underbrace{\left[ \begin{matrix}
  			{{I}_{m}} & {{D}_{p\,\,\text{reg}}}  \\
  			{{0}_{n\times m}} & {{B}_{p}}  \\
  		\end{matrix} \right]}_{B}\underbrace{\left( \left[ \begin{matrix}
  			-{{y}_{\text{cmd}}} \\ 
  			{u_{\text{bl}}}\left( t \right) \\ 
  		\end{matrix} \right]+\left[ \begin{matrix}
  			v\left( t \right) \\ 
  			w\left( t \right) \\ 
  		\end{matrix} \right] \right)}_{{{{\tilde{u}}}_{\text{bl}}}\left( t \right)+ \pi \left( t \right)} \\ 
  	\underbrace{\left[ \begin{matrix}
  			{{e}_{y\,I}}\left( t \right)  \\
  			{{y}_{p}}\left( t \right)  \\
  		\end{matrix} \right]}_{y\left( t \right)}=\underbrace{\left[ \begin{matrix}
  			{{I}_{m}} & {{0}_{m\times n}}\,  \\
  			{{0}_{{{n}_{y}}\times m}} & {{C}_{p}}  \\
  		\end{matrix} \right]}_{C}\underbrace{\left[ \begin{matrix}
  			{{e}_{y\,I}}\left( t \right)  \\
  			{{x}_{p}}\left( t \right)  \\
  		\end{matrix} \right]}_{x\left( t \right)}+\underbrace{\left[ \begin{matrix}
  			{{0}_{m\times m}}  \\
  			{{D}_{p}}  \\
  		\end{matrix} \right]}_{D}u\left( t \right). \\ 
  \end{matrix}
\end{aligned}
\end{equation}
  Second, the limited output constraints and control input constraints are written in terms of the extended system,
 \begin{equation}
 	\label{eq42}
 	\begin{aligned} 
 		{{y}_{\lim }}\left( t \right)=\left[ \begin{matrix}
  	{u_{\text{bl}}}\left( t \right)  \\
  	{{z}_{\lim }}\left( t \right)  \\
  \end{matrix} \right]=
   \begin{bmatrix}
  		-{{K}_{I}} & -{{K}_{P}}  \\
  		{{0}_{m\times m}} & {{C}_{p\,\lim }}\,  \\
  	\end{bmatrix} 
  \left[ \begin{matrix}
  	{{e}_{y\,I}}\left( t \right)  \\
  	{{x}_{p}}\left( t \right)  \\
  \end{matrix} \right]={{C}_{\lim }}\,x\left( t \right),\text{   }{{y}^{\max }}=\left[ \begin{matrix}
  	{u^{\max }}  \\
  	z_{\lim }^{\max }  \\
  \end{matrix} \right],\text{   }{{y}^{\min }}=\left[ \begin{matrix}
  	{u^{\min }}  \\
  	z_{\lim }^{\min }  \\
  \end{matrix} \right]
\end{aligned}
\end{equation}
where ${{K}_{I}}$ and ${{K}_{P}}$ are the proportional and integral gain matrices, respectively, and we define $K_x=[K_I\quad K_P]$.
  Since~\eqref{eq41} and~\eqref{eq42} are now expressed in the form~\eqref{eq1}, the baseline control with augmentation in~\eqref{eq39},~\ref{eq40} can be applied.

  \begin{assum}
  	\label{assum5}
  	For the PI servo-control design case, the constraints~\eqref{eq38} are feasible.
  \end{assum}

  \subsection{Constraint Satisfaction and Forward Invariance}
  
  The augmented system~\eqref{eq41} and constrained output~\eqref{eq42} are used to simultaneously enforce state and input constraints. 
  Importantly, in \cite{LavretskyMenner2025}, it is shown that limiting ${u_{\text{bl}}}\left( t \right)$in~\eqref{eq42} implies that the total control command $u\left( t \right)$evolves within the same bounds. 
  Theorem~\ref{thm4} with proof in the appendix shows that enforcing constraints for the baseline controller as in~\eqref{eq42} also enforces the same limit on the total control command and the integrator states~$e_{y\, I}$ remain bounded. Here, we assume that the constraints~\eqref{eq38} are feasible, which implies that a controller can be found that satisfies~\eqref{eq38}. Future work will focus on conditions for feasibility.
  
  \begin{thm}
  	\label{thm4}
  	Consider the controllable LTI system~\eqref{eq37} Suppose a stabilizing baseline control policy with the CBF augmentation as in~\eqref{eq3}. Then the following statements take place:
  	(i) Enforcing the operational constraints on the baseline control command in~\eqref{eq2} implies that the total control command is bounded with~\eqref{eq38}. 
  	(ii) The controller integrator state with the dynamics~\eqref{eq40} remains UUB forward in time.
  \end{thm}

  \subsection{Robustness Margins and Relative Stability}
  
  For this analysis, all active AW augmentation loops due to $v\left( t \right)$ need to be closed, since the integrator dynamics represent a known part of the total controller. Hence, we need to separate the AW signal $v\left( t \right)$ from the control augmentation signal $w\left( t \right)$ for robustness analysis, where we define
   \begin{equation}
  	\label{eq44} 
  	\begin{matrix} 
  		v\left( t \right) =-{{\delta }_{v}}\left( x\left( t \right) \right)\left( {{H}_{\pi}}
  		\begin{bmatrix}
  			{{0}_{m\times m}}  \\
  			{{I}_{m}} 
  		\end{bmatrix}{{K}_{x}}+{{H}_{x}} \right)x\left( t \right)
  \\
  w\left( t \right) =-{{\delta }_{w}}\left( x\left( t \right) \right)\left( {{H}_{\pi}}\begin{bmatrix}
  	{{0}_{m\times m}}  \\
  	{{I}_{m}} 
  \end{bmatrix} {{K}_{x}}-{{H}_{x}} \right)x\left( t \right)
  \\
  {{\delta }_{v}}\left( x\left( t \right) \right) = \begin{bmatrix}
  	{{I}_{m}} & {{0}_{m\times m}}  \\
  \end{bmatrix}H_{\pi}^{-1}\delta \left( x\left( t \right) \right)
  \\
   {{\delta }_{w}}\left( x\left( t \right) \right) =\left[ \begin{matrix}
  	{{0}_{m\times m}} & {{I}_{m}}  \\
  \end{matrix} \right]H_{\pi}^{-1}\delta \left( x\left( t \right) \right).
\end{matrix}
\end{equation} 
  The system loop gain transfer function matrix ${{L}_u}\left( s \right)$ can be computed with the total servo-controller ${u_{\text{out}}}\left( t \right)=-{{K}_{x}}x\left( t \right)+w(t)$, while zeroing out command terms ${{y}_{\text{cmd}}}$ and $y_{\lim }^{\min /\max }$.
  The closed-loop system dynamics yields
   \begin{equation}
  	\label{eq48}
  	\begin{matrix} 
  		\dot{x}\left( t \right)=A\,x\left( t \right)+B\begin{bmatrix}
  	v\left( t \right)  \\
  	u\left( t \right)  \\
  \end{bmatrix} =\left( A-
  {A}_{v}\left( {{\delta }_{v}} \right)
  \right)\,x\left( t \right)+B \begin{bmatrix}
  	{{0}_{m\times m}}  \\
  	{{I}_{m}}  \\
  \end{bmatrix} u\left( t \right)
  \\
  {A}_{v}\left( {{\delta }_{v}} \right)=B\left[ \begin{matrix}
  	{{I}_{m}}  \\
  	{{0}_{m\times m}}  \\
  \end{matrix} \right]{{\delta }_{v}}\left( {{H}_{\pi}}\left[ \begin{matrix}
  	{{0}_{m\times m}}  \\
  	{{I}_{m}}  \\
  \end{matrix} \right]{{K}_{x}}+{{H}_{x}} \right).
\end{matrix}
\end{equation} 
  Finally, we can state the loop gain ${{L}_{u}}\left( s;\,\delta  \right)$ using ${{u}_{\text{out}}}\left( t \right)=-{{L}_{u}}\left( s;\,\delta  \right){{u}_{\text{in}}}\left( t \right)$
   \begin{equation}
  	\label{eq50}
  	\begin{aligned} 
  {{L}_{u}}\left( s;\,\delta  \right)=\left( {{K}_{x}}-{{\delta }_{w}}\left( {{H}_{\pi}}\begin{bmatrix}
  	{{0}_{m\times m}}  \\
  	{{I}_{m}}  \\
  \end{bmatrix}{{K}_{x}}-{{H}_{x}} \right) \right){{\left( s\,{{I}_{n}}-\left( A-{{A}_{v}}\left( {{\delta }_{v}} \right) \right) \right)}^{-1}}B \begin{bmatrix}
  	{{0}_{m\times m}}  \\
  	{{I}_{m}}  \\
  \end{bmatrix}.
  \end{aligned}
\end{equation}
  In this case, SISO and MIMO margins at the system input breakpoint are defined based on the resulting $\left( m\times m \right)-$dimensional loop gain transfer function matrix, parameterized with the binary-valued matrix $\delta $.

  \section{Augmentation Design for Output Feedback Controllers}
  
  In practice, the full state is often not accessible and needs to be estimated.
  This section deals with the observer-based output feedback case and discussed key differences to the generic design in Section~\ref{sec3}, see~\cite{menner2026output} for further details.
  Consider the dynamical system in~\eqref{eq1} without full access to the state,
  \begin{equation}
  	\label{eq27}
  	\begin{aligned} \begin{matrix}
  			\dot{x}\left( t \right)=Ax\left( t \right)+Bu\left( t \right) \\ 
  			y\left( t \right)=Cx\left( t \right)+Du\left( t \right) \\ 
  			{{y}_{\lim }}\left( t \right)={{C}_{\lim }}\,x\left( t \right) \\ 
  		\end{matrix}
  	\end{aligned}
  \end{equation}
  where $y\left( t \right)\in {{R}^{{{n}_{y}}}}$ is the vector of measured outputs. Note that $C$ and ${{C}_{\lim }}$ do not have to be identical. The method in this paper can limit a state that is not directly measurable.

  Let the control command in~\eqref{eq3} use the baseline controller
  \begin{equation}
  	\label{eq28}
  	\begin{aligned} 
  		{{u}_{\text{bl}}}\left( t \right)=-K\hat{x}\left( t \right),
  	\end{aligned}
  \end{equation}
  feedback gain matrix $K$, and the state estimate $\hat{x}\left( t \right)$ provided by the observer with dynamics
  \begin{equation}
  	\label{eq29}
  	\begin{aligned} 
  		\begin{matrix}
  			\dot{\hat{x}}\left( t \right)=A\hat{x}\left( t \right)+Bu\left( t \right)+L\left( y\left( t \right)-\hat{y}\left( t \right) \right) \\ 
  			\hat{y}\left( t \right)=C\hat{x}\left( t \right)+Du\left( t \right) \\ 
  		\end{matrix}
  	\end{aligned}
  \end{equation}
  with Luenberger gain $L$ and output estimate $\hat{y}\left( t \right)$ \cite{Luenberger1964,Luenberger1966,Khalil2001}. Both the feedback gain matrix and the Luenberger gain can, e.g., be designed using a linear quadratic regulator (LQR) design. For the output feedback case, the control augmentation signal in~\eqref{eq13} needs to be expressed in terms of the state estimate rather than the state,
  \begin{equation}
  	\label{eq30}
  	\begin{aligned} 
  		& \pi \left( \hat{x}\left( t \right) \right)=H_{\pi }^{-1}\left( \max \left( {{0}_{m\times 1}},\ \Delta {{H}_{\min }}\left( \hat{x}\left( t \right),\,\,{{u}_{\text{bl}}}\left( t \right) \right) \right)-\max \left( {{0}_{m\times 1}},\ \Delta {{H}_{\max }}\left( \hat{x}\left( t \right),\,\,{{u}_{\text{bl}}}\left( t \right) \right) \right) \right) \\ 
  		& =
  		H_{\pi }^{-1}
  		\left\{ 
  		\begin{matrix}
  			& \left( -{{H}_{x}}\hat{x}\left( t \right)-{{H}_{\pi }}{{u}_{\text{bl}}}\left( t \right)+{{\alpha }_{\pi }}y_{\lim }^{\min } \right) & \text{if}\ \left( -{{H}_{x}}\hat{x}\left( t \right)-{{H}_{\pi }}{{u}_{\text{bl}}}\left( t \right)+{{\alpha }_{\pi }}y_{\lim }^{\min } \right)>0 
  			\\ 
  			& -\left( {{H}_{x}}\hat{x}\left( t \right)+{{H}_{\pi }}{{u}_{\text{bl}}}\left( t \right)-{{\alpha }_{\pi }}y_{\lim }^{\max } \right) & \text{if}\ \left( {{H}_{x}}\hat{x}\left( t \right)+{{H}_{\pi }}{{u}_{\text{bl}}}\left( t \right)-{{\alpha }_{\pi }}y_{\lim }^{\max } \right)>0 \\ 
  			& 0 & \text{otherwise} \\ 
  		\end{matrix} 
  		\right.  
  	\end{aligned}
  \end{equation}
  The results in Section~\ref{sec3} can directly be used for CBF-ability and closed-loop stability~\cite{menner2026output}.
  However, constraint satisfaction requires a more in-depth analysis.
  It is clear that $\pi (\hat{x}(t))=\pi (x(t))$ when the state estimation error has converged, ${{e}_{\text{est}}}\left( t \right)=x\left( t \right)-\hat{x}\left( t \right)=0.$ 
  However, during transients with ${{e}_{\text{est}}}(t)\ne 0$, compared to the state feedback case in~\eqref{eq13}, the operational constraints are not always satisfied by construction for any choice of design parameters. 
  Instead, for the output feedback-based control augmentation policy~\eqref{eq30}, attention needs to be paid to selecting the design parameters appropriately, which is discussed in the following.

  \begin{rem}[CBF-ability and Closed-Loop Stability]
  	Akin to the modified constrained output ${{Y}_{\lim }}\left( t \right)$ in~\eqref{eq10}, we can define the constrained limited output using the state estimate ${{\hat{Y}}_{\lim }}\left( t \right)={{H}_{x}}\hat{x}\left( t \right)+{{H}_{\pi }}u\left( t \right)$.
  	Then, CBF-ability and closed loop stability can be shown using similar argument as in Section~\ref{sec3} and the separation principle, i.e., the error dynamics are independent of the baseline control and CBF augmentation,
  	\begin{equation}
  		\label{eq33}
  		\begin{aligned} 
  			\begin{bmatrix}
  				\dot{x}\left( t \right) \\ 
  				{{{\dot{e}}}_{\text{est}}}\left( t \right) \\ 
  			\end{bmatrix} = \begin{bmatrix}
  				A-BH_{\pi }^{-1}{{H}_{x}} & BH_{\pi }^{-1}{{H}_{x}}  \\
  				0 & A-LC  \\
  			\end{bmatrix} \begin{bmatrix}
  				x\left( t \right) \\ 
  				{{e}_{\text{est}}}\left( t \right) \\ 
  			\end{bmatrix} + \begin{bmatrix}
  				BH_{\pi }^{-1}{{{\hat{Y}}}_{\lim }}\left( t \right) \\ 
  				0 \\ 
  			\end{bmatrix}.
  		\end{aligned}
  	\end{equation}
  \end{rem}

  \begin{assum}
  	For the output feedback case~\eqref{eq27}, the matrix pair $\left( A,C \right)$ is observable.
  \end{assum}

  		\subsection{Constraint Satisfaction and Forward Invariance}
  		
  		To show constraint satisfaction, we use exponential decay/stability arguments of the state estimation error dynamics and of the operational constraint, i.e., when the modified constraint $H_i$ is active, the operational constraint $h_i=0$ is being approached with exponential decay.
  		We present a sufficient condition for the state – not only the state estimate – satisfying the operational boundaries. Consider constraint i with relative degree $r_i$ and CBF parameters {$\alpha_{i0},\alpha_{i1},...,\alpha_{i(r_i-1)}$}, where $\alpha_{ij}=-\lambda_{ij}$ defining the eigenvalues of the stable polynomial~\eqref{eq9}. Theorem~\ref{thm3} with proof in the appendix shows that if at least one CBF parameter $\alpha_{ij}$ is chosen to be smaller than the slowest dynamics of the observer, 
  		\begin{equation}
  			\label{eq34}
  			\begin{aligned} 
  				\begin{matrix}
  					\alpha _{i}^{\star }<\left| \lambda _{\max }^{\text{Re}} \right| \\ 
  					\alpha _{i}^{\star }=\min \{{{\alpha }_{i0}},{{\alpha }_{i1}},...{{\alpha }_{i({{r}_{i}}-1)}}\} 
  				\end{matrix}
  			\end{aligned}
  		\end{equation}
  		where $\lambda _{\max }^{\text{Re}}={{\max }_{j}}\operatorname{Re}\left( {{\lambda }_{j}}\left( A-LC \right) \right)$ is the maximum (slowest) eigenvalue of the observer, then there exists a finite time ${{t}_{h\le 0}}$ such that forward invariance is guaranteed forward in time, $h\left( x\left( t \right) \right)\le 0$ for all $t\ge {{t}_{h\le 0}}$. Theorem~\ref{thm3} implies that if the operational constraint is satisfied at any time ${{t}_{0}}$ during operation, it is guaranteed to be satisfied for all $t\ge {{t}_{0}}+{{t}_{h\le 0}}$, i.e., after an initial transient in the estimation error has diminished. 
  		Assuming the initial estimation error to be sufficiently small, we can show that ${{t}_{0}}={{t}_{h\le 0}}$. 
  		
  		\begin{thm}
  			\label{thm3}
  			Consider the output feedback augmentation~\eqref{eq30}. Let the CBF parameters be chosen as in~\eqref{eq34}. For any initial condition $x\left( t=0 \right)$ with $h\left( x\left( t=0 \right) \right)\le 0$, there exists a finite time ${{t}_{h\le 0}}$ such that $h\left( x\left( t \right) \right)\le 0$ for all $t\ge {{t}_{h\le 0}}$.
  		\end{thm}

  		\subsection{Robustness Margins and Relative Stability}
  		The key difference of the output feedback case compared to state feedback is the additional observer dynamics. Using ${{K}_{i,\text{CBF}}}=H_{\pi }^{-1}\delta \left( \hat{x}\left( t \right) \right)\left( {{H}_{x}}-{{H}_{\pi }}K \right)$, the state estimate, the transfer function at the input breakpoint, and loop gain are given by
  		\begin{equation}
  			\label{eq35}
  			\begin{aligned} 
  				\begin{matrix}
  					\hat{x}\left( t \right)={{\left( s\,{{I}_{n}}-\left( A-B\left( K+{{K}_{i,\text{CBF}}} \right)-LC \right) \right)}^{-1}}LCx\left( t \right) \\ 
  					{{u}_{\text{out}}}\left( t \right)=-{{L}_{u}}\left( s;\,\delta  \right){{u}_{\text{in}}}\left( t \right) 
  					\\
  					{{L}_{u}}\left( s;\,\delta  \right)=\left( K+{{K}_{i,\text{CBF}}} \right){{\left( s\,{{I}_{n}}-\left( A-B\left( K+{{K}_{i,\text{CBF}}} \right)-LC \right) \right)}^{-1}}LC{{\left( s\,{{I}_{n}}-A \right)}^{-1}}B.
  				\end{matrix}
  			\end{aligned}
  		\end{equation}

  \section{Trade Studies}

  \subsection{Short Period Dynamics with Output Feedback and PI Control with Gust Input  }
  This section applies the proposed output feedback control method to a flight control example using PI control with command tracking. Consider the short-period dynamics (see \cite{Lavretsky2024}, Section 14.7),
  \begin{equation}
  	\label{eq55}
  	\begin{aligned} 
  		\begin{matrix}
  	\left[ \begin{matrix}
  		{{{\dot{e}}}_{I,{{C}^{\star }}}}\left( t \right) \\ 
  		\dot{\alpha }\left( t \right) \\ 
  		\dot{q}\left( t \right) \\ 
  	\end{matrix} \right]=\left[ \begin{matrix}
  		0 & \text{6}\text{.78} & \text{1}\text{.29}  \\
  		0 & -2.24 & 0.990  \\
  		0 & -4.47 & -0.902  \\
  	\end{matrix} \right]\left[ \begin{matrix}
  		{{e}_{I,{{C}^{\star }}}}\left( t \right) \\ 
  		\alpha \left( t \right) \\ 
  		q\left( t \right) \\ 
  	\end{matrix} \right]+\left[ \begin{matrix}
  		\text{0}\text{.876} \\ 
  		-0.233 \\ 
  		-4.59 \\ 
  	\end{matrix} \right]{{\delta }_{\text{ele}}}\left( t \right)+\left[ \begin{matrix}
  		-1 \\ 
  		0 \\ 
  		0 \\ 
  	\end{matrix} \right]C_{\text{cmd}}^{\star }+\left[ \begin{matrix}
  		0  \\
  		-2.24  \\
  		-4.47  \\
  	\end{matrix} \right]\delta {{w}_{G}}\left( t \right) \\ 
  	{{\delta }_{\text{ele}}}\left( t \right)={{\delta }_{\text{ele},\text{bl}}}\left( t \right)+w\left( t \right) \\ 
  	C_{\text{cmd}}^{\star }=C_{\text{cmd,bl}}^{\star }-\nu \left( t \right) \\ 
  \end{matrix}
\end{aligned}
\end{equation}
  with the angle of attack $\alpha \left( t \right)$, pitch rate $q\left( t \right)$, and integral regulated output error ${{e}_{I,{{C}^{\star }}}}\left( t \right)$. The regulated output is ${{C}^{\star }}\left( t \right)$, which is a linear combination of vertical load factor ${{N}_{z}}\text{  }(g)$ and pitch rate. The control input is ${{\delta }_{\text{ele}}}\left( t \right)$ and the command is $C_{\text{cmd}}^{\star }$. Additionally, we consider a disturbance/gust input $\delta {{w}_{G}}\left( t \right)\text{  }(\text{rad})$ impacting the closed-loop system as an angle of attack disturbance. The measured and limited outputs are
  \begin{equation}
  	\label{eq56}
  	\begin{aligned} 
  		\begin{matrix}
  	\text{Measured Output:   }y\left( t \right)={{\left[ \begin{matrix}
  				{{e}_{I,{{C}^{\star }}}}\left( t \right) & q\left( t \right)  \\
  			\end{matrix} \right]}^{T}}=C{{\left[ \begin{matrix}
  				{{e}_{I,{{C}^{\star }}}}\left( t \right) & \alpha \left( t \right) & q\left( t \right)  \\
  			\end{matrix} \right]}^{T}}  \\
  		\alpha _{\lim }^{\min } \le \alpha \left( t \right)\le \alpha _{\lim }^{\max }  \\ 
  	{{\delta }_{\text{ele}}}_{{}}^{\min } \le {{\delta }_{\text{ele}}}\left( t \right)\le {{\delta }_{\text{ele}}}_{{}}^{\max }, 
  \end{matrix}
\end{aligned}
\end{equation}
where the bounds are varied to show different constraint satisfaction behaviors and highlight properties.
  The controller design uses an LQR for both the feedback gain and Luenberger gain resulting in
  \begin{equation}
  	\label{eq58}
  	\begin{aligned} 
  		K=\left[ \begin{matrix}
  	-0.310 & -0.251 & -0.399  \\
  \end{matrix} \right],L=\left[ \begin{matrix}
  	\begin{matrix}
  		3.16  \\
  		2.69  \\
  		0.947  \\
  	\end{matrix} & \begin{matrix}
  		9.29  \\
  		10.85  \\
  		5.92  \\
  	\end{matrix}  \\
  \end{matrix} \right]
\end{aligned}
\end{equation}
  The resulting eigenvalues are
  $ \text{eig}\left( A-BK \right)=\{-1.52,-\text{1}\text{.62}+\text{2}\text{.30i},-\text{1}\text{.62}-\text{2}\text{.30i}\}$ and  $\text{eig}\left( A-LC \right)=\{-\text{3}\text{.28},-\text{3}\text{.05}+\text{11}\text{.6i},-\text{3}\text{.05}-\text{11}\text{.6i}\}$.
  As supported by our theoretical derivations, we choose the following CBF tuning parameter ${{\alpha }_{\text{ele}}}=2$ and ${{\alpha }_{\text{AOA}}}=1.5.$
 We use a gust input/disturbances modeled as in~\cite{Lavretsky2024} (Fig. 14.8 Gust dynamic model, p. 608). 
 The initial state estimate and initial state are 
 $
  	\hat{x}\left( t=0 \right)=0$, $x\left( t=0 \right)={{\left[ 0 \quad -4\deg  \quad 0  \right]}^{T}}$, and ${{e}_{\text{est}}}\left( t=0 \right)={{\left[0 \quad -4\deg  \quad 0 \right]}^{T}}$. 
The commanded value can be seen in Fig.~\ref{figAW_bl}.

\subsubsection{Input Constraint and Anti-Windup Protection}
\begin{figure}[t] 
	\centerline{ 
		\includegraphics[width=7.5cm]{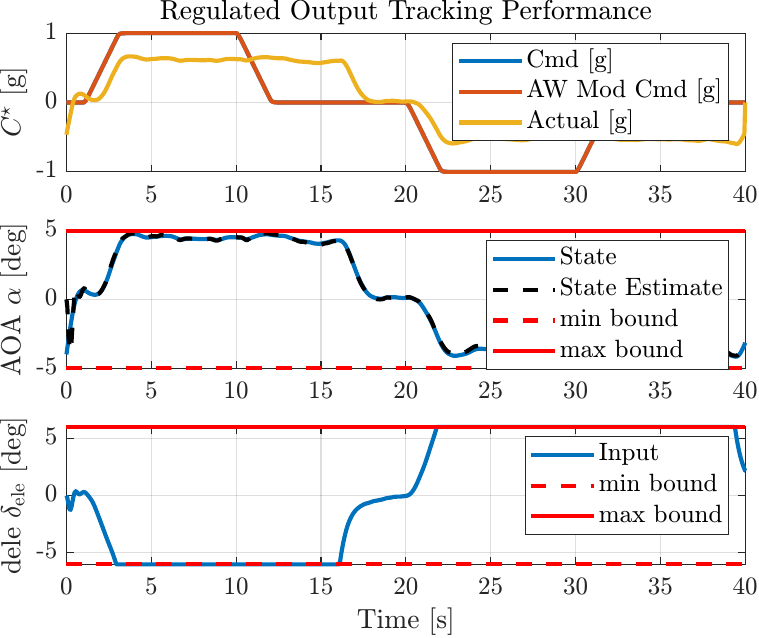}
		\includegraphics[width=7.5cm]{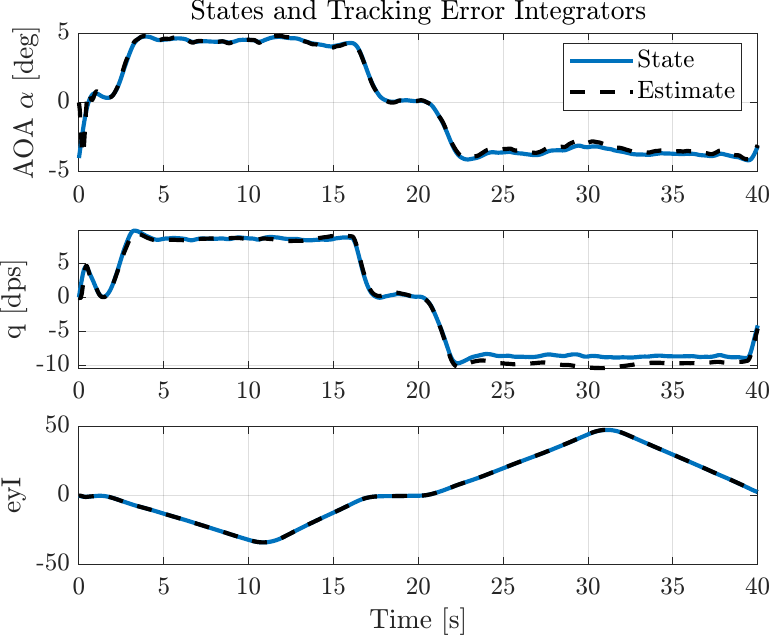} 
	} 
	\caption{Anti-windup example, baseline LQR PI controller and saturation logic. 
		Left: Tracking performance and operational limits. 
		Right: State and state estimate.} 
	\label{figAW_bl} 
\end{figure}

Let $\alpha _{\lim }^{\min }=-5\deg,\alpha _{\lim }^{\max }=5\deg$ and ${{\delta }_{\text{ele}}}_{{}}^{\min }=-6\deg, {{\delta }_{\text{ele}}}_{{}}^{\max }=6\deg  $ to highlight input constraint and integrator windup behavior.
Fig.~\ref{figAW_bl} shows a scenario without the CBF augmentation but a hard saturation logic to clip the elevator input to its max/min values.
Clearly, this forces input constraint satisfaction. However, since the hard saturation logic does not change the commanded and unreachable $C^\star$ value, the integrator state winds up. 
As a consequence, the commanded $C^{\star}$ tracking is delayed as the integrator state needs to unwind, see $C^\star$ in 10-16s in Fig.~\ref{figAW_bl}.

\begin{figure}[t] 
\centerline{ 
	\includegraphics[width=7.5cm]{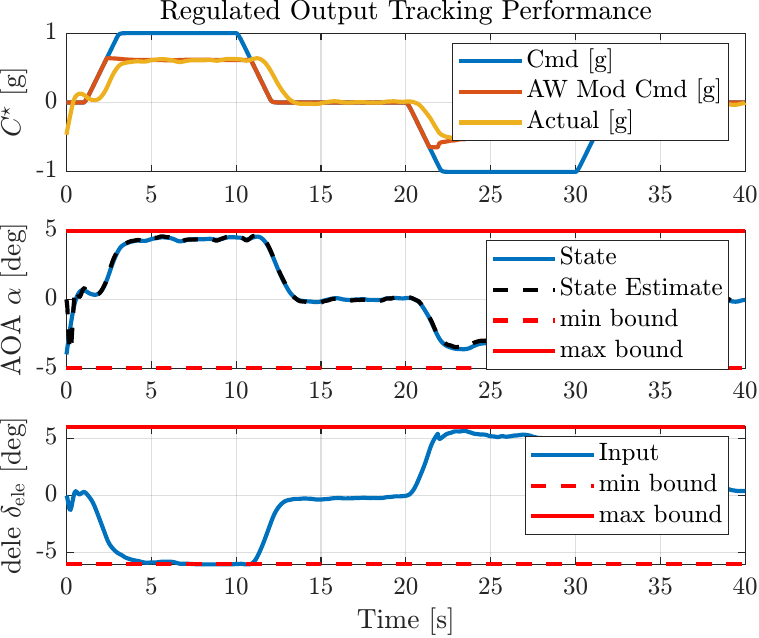}
	\includegraphics[width=7.5cm]{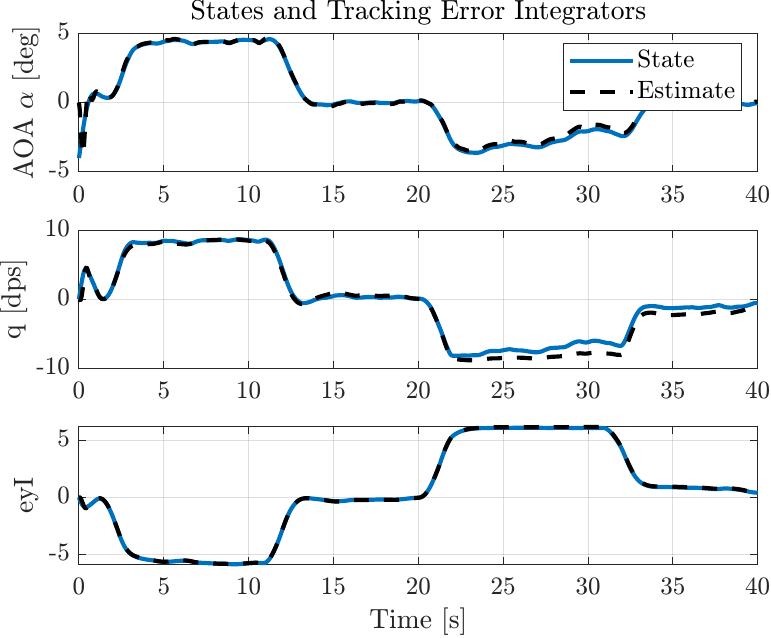} 
} 
\caption{Anti-windup example, baseline LQR PI controller and CBF augmentation. 
	Left: Tracking performance and operational limits. 
	Right: State and state estimate.} 
\label{figAW_cbf} 
\end{figure}

The proposed augmentation provides a systematic anti-windup protection and keeps to integral error state bounded by construction. Fig.~\ref{figAW_cbf} shows that the proposed augmentation modifies the commanded $C^\star$ to a value that is physically possible to track. As a consequence, input constraints are satisfied without the need for a hard saturation logic and the integral error state remains bounded.

\subsubsection{State Constraint}
\begin{figure}[t] 
	\centerline{ 
		\includegraphics[width=7.5cm]{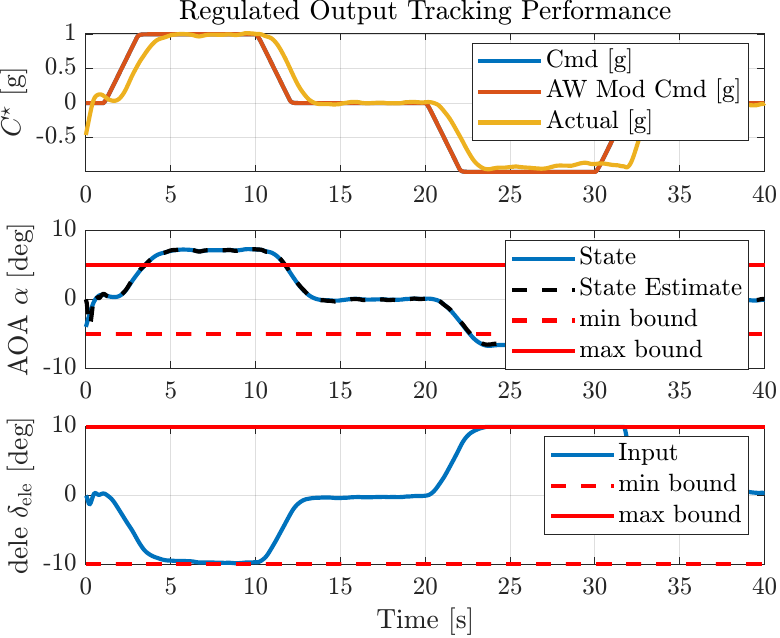}
		\includegraphics[width=7.5cm]{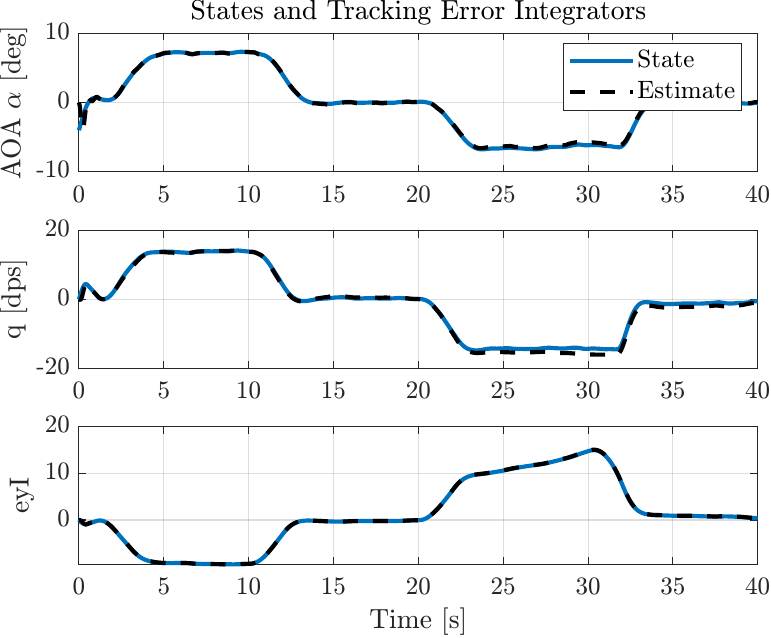} 
	} 
	\caption{State limiting example, baseline LQR PI controller and saturation logic. 
		Left: Tracking performance and operational limits. 
		Right: State and state estimate.} 
	\label{figAOA_bl} 
\end{figure}

Let $\alpha _{\lim }^{\min }=-5\deg,\alpha _{\lim }^{\max }=5\deg$ and $
{{\delta }_{\text{ele}}}_{{}}^{\min }=-10\deg, {{\delta }_{\text{ele}}}_{{}}^{\max }=10\deg  $. 
The baseline controller with hard input saturation logic is not capable of protecting against state constraint violations, see Fig.~\ref{figAOA_bl}. Instead, the baseline controller violates the angle of attack constraint in order to track $C^\star$.

\begin{figure}[t] 
	\centerline{ 
		\includegraphics[width=7.5cm]{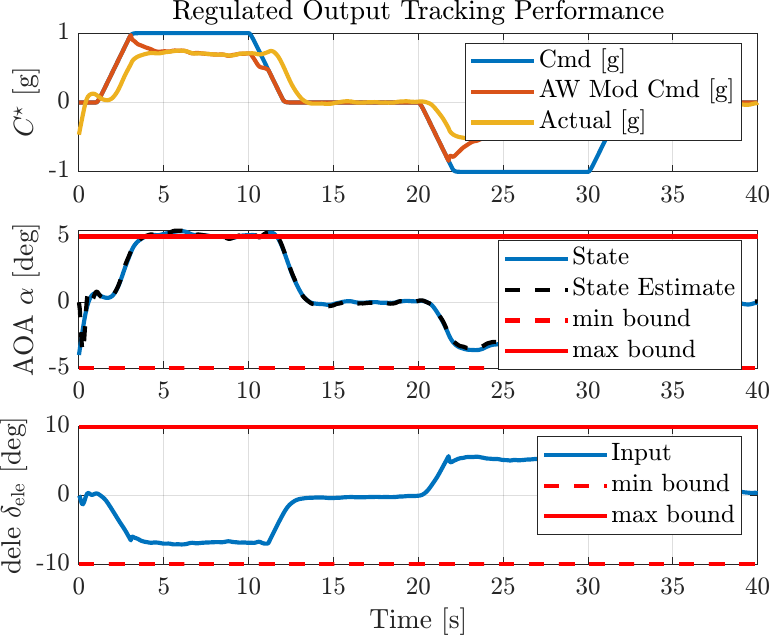}
		\includegraphics[width=7.5cm]{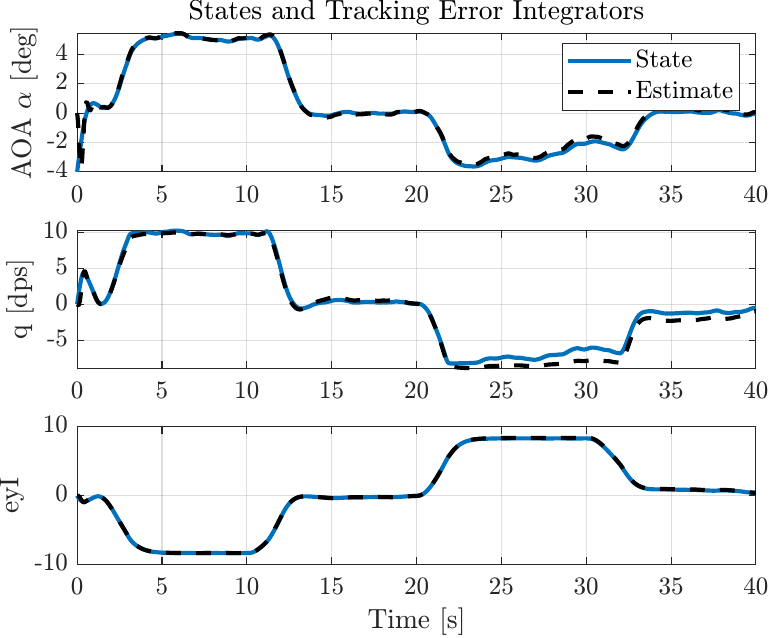} 
	} 
	\caption{State limiting example, baseline LQR PI controller and CBF augmentation. 
		Left: Tracking performance and operational limits. 
		Right: State and state estimate.} 
	\label{figAOA_cbf} 
\end{figure}

In contrast, our proposed method offers a systematic protection against both state and input constraint violations. 
Here, it can be seen that the $C^\star$ tracking is sacrificed in order to ensure state constraints are respected, see anti-windup modified command at 3-10s, see Fig.~\ref{figAOA_cbf}.
Due to the gust/disturbance signal, the proposed control augmentation cannot guarantee constraint satisfaction for all times, which is no different from any other constrained control design. What is different in the proposed method, these temporary constraint violations do not cause infeasibilities since we use a QP only for design – not during execution. In practice, the limits enforced by the augmentation should be chosen sufficiently far away from the aircraft’s physical limits in order to prevent physical limits from being exceeded. 
The proposed augmentation remains closed-loop even in the presence of active constraints, which implies a tracking behavior of the limits. This design feature increases robustness, which is highlighted next. 


\subsubsection{Robustness Margins}

One significant advantage of the proposed CBF augmentation design is its analyzability through linear systems theory. To illustrate this, Table~\ref{tab2} presents MIMO margins computed at the input breakpoint. It shows the nominal gain and phase margins using the model in~\eqref{eq55} as well as a scenario with second order (unmodeled) actuator dynamics, 
${{\ddot{\delta }}_{\text{ele}}}\left( t \right)=-98{{\dot{\delta }}_{\text{ele}}}\left( t \right)+4900\left( {{\delta }_{\text{ele,cmd}}}-{{\delta }_{\text{ele}}}\left( t \right) \right).$
In the absence of active constraints, the proposed method is identical to the baseline control scenario. 
When the elevator limit is reached, the proposed method with active aileron constraint remains closed-loop and exhibits comparable margins to the unconstrained case, while ensuring constraint satisfaction, see second row. 
This analysis highlights the robustness and effectiveness of the proposed augmentation in managing constraints while preserving system stability.

\begin{table}[t]
	\begin{center}
		\begin{tabular}{c|c c} 
			\toprule
			\textbf{Active Constraints}	& \textbf{Baseline Controller} &	\textbf{Baseline with Augmentation} \\
			\midrule
			None & GM = $\infty$, PM = 83.5deg & GM = $\infty$, PM = 83.5deg \\
			None (with actuator dyn.) & GM = 25.8dB, PM = 81.9deg & GM = 25.8dB, PM = 81.9deg \\
			\midrule
			Elevator & 	N/A (constraints violated) & GM = $\infty$, PM = 102deg\\
			Elevator  (with actuator dyn.) & 	N/A (constraints violated) & 	GM = 30.4dB, PM = 101deg\\
			\midrule
			AOA & 	N/A (constraints violated) & GM = $\infty$, PM = 129deg\\
			AOA  (with actuator dyn.) & 	N/A (constraints violated) & 	GM = 48.0dB, PM = 113deg\\
			\midrule
			AOA, Elevator  & 	N/A (constraints violated) & GM = $\infty$, PM = 131deg\\
			AOA, Elevator  (with actuator dyn.) & 	N/A (constraints violated) & 	GM = 48.8dB, PM = 116deg\\
			\bottomrule
		\end{tabular}
		\caption{Multi-Input Multi-Output (MIMO) margins with and without constraint augmentation. 
			}
		\label{tab2}
	\end{center}
\end{table}

\subsection{Nonlinear Simulation Study}
\begin{figure}[t] 
	\centerline{ 
		\includegraphics[width=7.5cm]{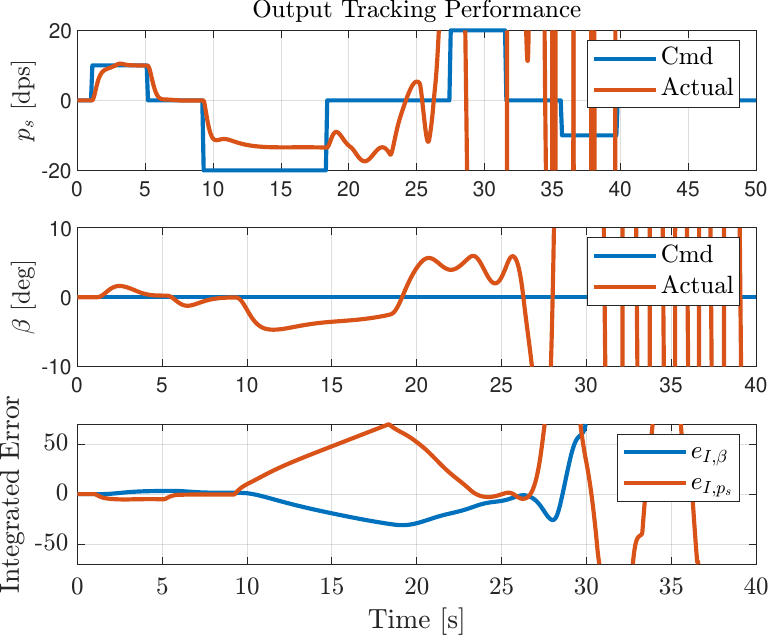}
		\includegraphics[width=7.5cm]{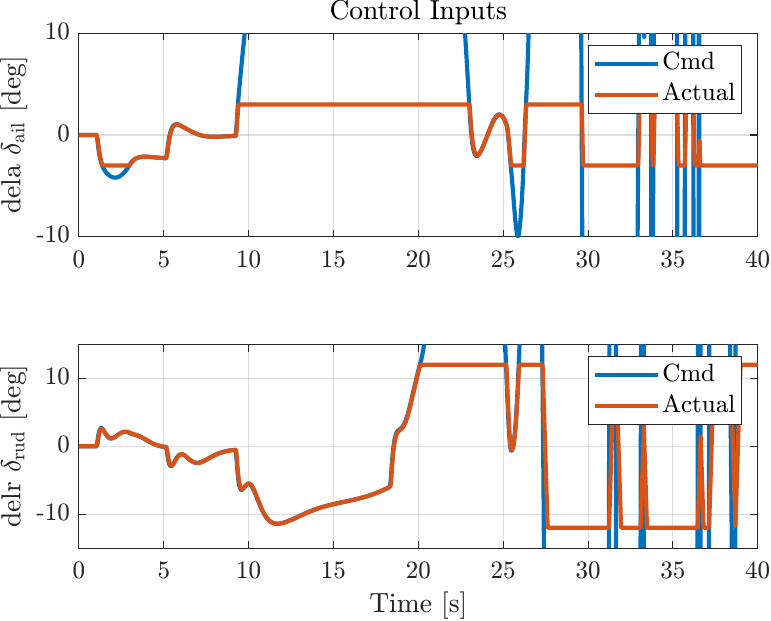} 
	} 
	\caption{Six-degree-of-freedom (6DoF) rigid body flight simulation results without proposed augmentation.} 
	\label{fig6dof_bl} 
\end{figure}

\begin{figure}[t] 
\centerline{ 
	\includegraphics[width=7.5cm]{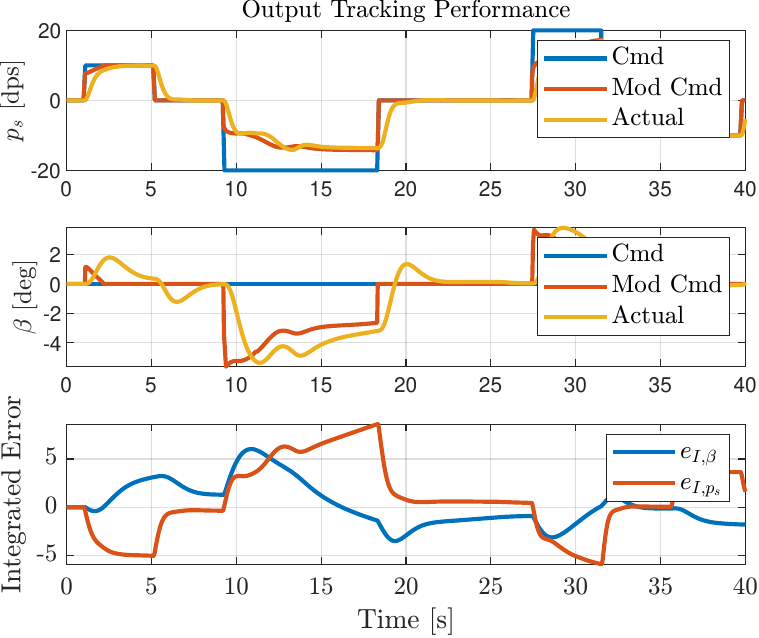}
	\includegraphics[width=7.5cm]{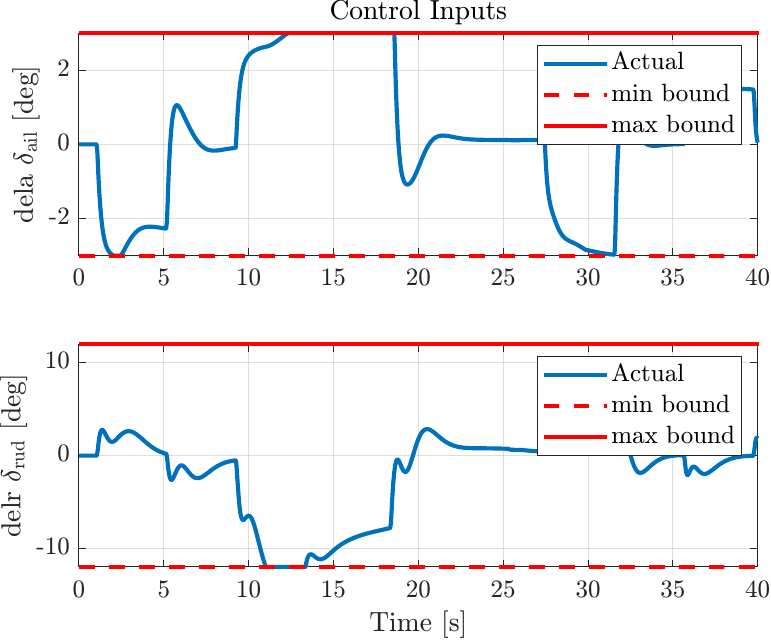} 
} 
\caption{Six-degree-of-freedom (6DoF) rigid body flight simulation results without proposed augmentation.} 
\label{fig6dof_cbf} 
\end{figure}

Next, we present simulation results derived using the six-degree-of-freedom (6DoF) rigid body flight simulation environment detailed in~\cite{Lavretsky2024}, Section~1.4.
The aircraft under study is representative of the McDonnell Douglas F-4 Phantom II and incorporates actuator dynamics, aerodynamic modeling, atmospheric conditions, and engine performance. 
The 6DoF model includes an airspeed controller, a pitch controller, and a roll-yaw controller. 
The pitch and roll-yaw controller utilize a proportional-integral feedback baseline controller augmented by the proposed CBF methodology.
Here, we focus on the AW component of the proposed augmentation and show roll-yaw states and control inputs.
In particular, we command the aircraft to follow a given modulated roll rate and maintain zero sideslip angle, see Fig.~\ref{fig6dof_bl} for the performance without the proposed augmentation but with a hard saturation logic on the control inputs and Fig.~\ref{fig6dof_cbf} with the proposed augmentation.

Initially, when a roll rate of ten degrees per second (dps) is commanded, the system operates without reaching any control input limits. As a result, the commanded roll rate is approached rapidly and tracked robustly, see the top left plot from 1 to 5 seconds, for both with and without augmentation. 
However, at 9 seconds, a roll rate command of -20dps is issued. 
In this scenario, the presence of aileron and rudder actuation limitations leads to a noticeable difference in transient performance compared to the unconstrained case. 
The commanded roll rate cannot be physically tracked due to the imposed aileron and rudder limits of $\pm 3$deg and $\pm 11$deg, respectively.

For the case without augmentation in Fig.~\ref{fig6dof_bl}, this inability to track the commanded roll rate leads to a constant tracking error $e_y=p_s-p_{s,{\rm cmd}}$ and thus to integrator windup.
The windup leads to a destabilization of the aircraft  before the integrator could wind down.
Using the proposed augmentation in Fig.~\ref{fig6dof_cbf}, the control limits are respected while error states are bounded. This is achieved by limiting/modifying the commanded roll rate to physically tractable values. 
Moreover, while the sideslip angle command remains zero throughout the simulation, the AW-modified sideslip
command is nonzero. 
This adjustment arises from the necessity to compensate for the limitations of the aileron and
rudder actuation. 
To effectively track the roll rate while operating under these constraints, the system augments the
command with a small sideslip angle. 
This modification exemplifies how the proposed CBF augmentation enhances the controller's ability to maintain performance and stability in the presence of operational constraints, ultimately ensuring that the aircraft can respond to external commands as effectively as possible.


  \section{Conclusions}
   This paper has developed a systematic framework for designing servo-controllers that explicitly enforce operational constraints on MIMO LTI dynamical systems. Leveraging foundational results from the Nagumo Theorem, the Comparison Lemma, and principles of min-norm optimal control, the methodology constructs a control augmentation that guarantees forward invariance of state and output trajectories within prescribed box constraints. Importantly, the control augmentation admits a closed-form, continuous piecewise-linear feedback law that enables rigorous closed-loop stability and robustness analyses grounded in linear system theory.
   The approach seamlessly extends to observer-based output feedback and PI control architectures, providing conditions to handle estimation errors and integrator windup while maintaining constraint satisfaction. Robustness margins computed at the control input breakpoint demonstrate that the augmented controller preserves favorable gain and phase margins even when constraints are active.
   Simulation studies spanning simple double integrator models, flight control short-period dynamics with gust disturbances, and six-degree-of-freedom nonlinear aircraft models confirm the performance and robustness of the proposed control augmentation in managing operational limits, improving transient behavior, and avoiding integrator windup without resorting to ad hoc saturation logic or online optimization.

  \section*{Appendix A: Proofs of Lemmas}

  \begin{proof}[Proof of Lemma~\ref{lem1acc}]  
  	 Applying ${{\phi }_{i}}(s)$ to the ${{i}^{\text{th}}}$ components of ${{h}_{\min}}(x)$ and ${{h}_{\max}}(x)$ gives
  \begin{equation}
  	\label{eq47acc}
  	\begin{aligned} 
  	{{H}_{\min\,i}}\left( x,\,{{u}_{\text{bl}}},\,\pi  \right)
  	&=\sum\limits_{j=0}^{{{r}_{i}}}{{{c}_{i\,j}}\left( y_{\lim }^{\min }-{{y}_{\lim }} \right)_{i}^{\left( j \right)}}={{c}_{i\,0}}{{\left( y_{\lim }^{\min } \right)}_{i}}-{{\left( {{Y}_{\lim }} \right)}_{i}} \\ 
  	{{H}_{\max\,i}}\left( x,\,{{u}_{\text{bl}}},\,\pi  \right)
  	&=\sum\limits_{j=0}^{{{r}_{i}}}{{{c}_{i\,j}}\left( {{y}_{\lim }}-y_{\lim }^{\max } \right)_{i}^{\left( j \right)}}=-{{c}_{i\,0}}{{\left( y_{\lim }^{\max } \right)}_{i}}+{{\left( {{Y}_{\lim }} \right)}_{i}} \\ 
  	{{\left( {{Y}_{\lim }} \right)}_{i}}
  	&=\sum\limits_{j=0}^{{{r}_{i}}}{{{c}_{i\,j}}\left( {{y}_{\lim }} \right)_{i}^{\left( j \right)}}={{c}_{i\,{{r}_{i}}}}\left( {{y}_{\lim }} \right)_{i}^{\left( {{r}_{i}} \right)}+\sum\limits_{j=1}^{{{r}_{i}}-1}{{{c}_{i\,j}}\left( {{y}_{\lim }} \right)_{i}^{\left( j \right)}} \\ 
  	& =\left( 
  	{{\left( {{C}_{\lim }} \right)}_{i}}A_{}^{{{r}_{i}}}\,x+{{\left( {{C}_{\lim }} \right)}_{i}}A_{}^{{{r}_{i}}-1}B\,\left(u_{\rm bl}+\pi\right)  
  	\right)+{{\left( {{C}_{\lim }} \right)}_{i}}\left( \sum\limits_{j=1}^{{{r}_{i}}-1}{{{c}_{i\,j}}A_{}^{j}} \right)x \\ 
  	& =\left( {{\left( {{C}_{\lim }} \right)}_{i}}A_{}^{{{r}_{i}}-1}B \right)\,\left(u_{\rm bl}+\pi\right) +{{\left( {{C}_{\lim }} \right)}_{i}}\left( \sum\limits_{j=0}^{{{r}_{i}}}{{{c}_{i\,j}}A_{}^{j}} \right)x, 
  	\end{aligned}
  \end{equation}
  with ${c}_{i\,{{r}_{i}}}=1$ for all $i$. Thus, ${{({{Y}_{\lim }})}_{i}}$ can be computed using the system parameters. The corresponding matrices defined by their ${{i}^{\text{th}}}$ rows are 
  \begin{equation}
  	\label{eq48acc}
  	\begin{aligned} 
  	& {{\left( {{H}_{\pi }} \right)}_{i}}={{\left( {{C}_{\lim }} \right)}_{i}}A_{}^{{{r}_{i}}-1}B \\ 
  	& {{\left( {{H}_{x}} \right)}_{i}}={{\left( {{C}_{\lim }} \right)}_{i}}\sum\limits_{j=0}^{{{r}_{i}}}{{{c}_{i\,j}}A_{}^{j}}={{\left( {{C}_{\lim }} \right)}_{i}}\prod\limits_{j=1}^{{{r}_{i}}}{\left( {{A}_{}}-{{\lambda }_{i\,j}}{{I}_{n}} \right)}.  
  	\end{aligned}
  \end{equation}
  Thus, ${{H}_{\min\,i}}\left( x,\,{{u}_{\text{bl}}},\,\pi  \right)=-{{\left( {{H}_{\pi }} \right)}_{i}}\pi -{{\left( {{H}_{x}} \right)}_{i}}x+{{c}_{i\,0}}{{\left( y_{\lim }^{\min } \right)}_{i}}$ and ${{H}_{\max\,i}}\left( x,\,{{u}_{\text{bl}}},\,\pi  \right)={{\left( {{H}_{\pi }} \right)}_{i}}\pi +{{\left( {{H}_{x}} \right)}_{i}}x-{{c}_{i\,0}}{{\left( y_{\lim }^{\max } \right)}_{i}}.$
\end{proof}

  \begin{proof}[Proof of Lemma~\ref{lem2}] 
  	By the Nagumo Theorem, the set $H$ is forward invariant for a dynamical system $\dot{x}=f(x)$ if, and only if, the state derivative vector $f(x)$ points inside the set at the boundary, i.e.,
  	\begin{equation}
  		\label{eq65}
  		\begin{aligned} 
  			\left[ x\left( t \right)\in H,\ \forall t\ge 0 \right]\Leftrightarrow \left[ {{\left. \dot{h}\left( x \right) \right|}_{h\left( x \right)=0}}\le 0,\ \forall t\ge 0 \right]\text{. }
  		\end{aligned}
  	\end{equation}
  	For $r=1$, if ${{\tilde{H}}_{1}}$ is forward invariant, then  ${{\tilde{h}}_{1}}\left( x \right)={{\phi }_{1}}\left( s \right)h\left( x \right)=\left( s-{{\lambda }_{1}} \right)h\left( x \right)=\dot{h}\left( x \right)-{{\lambda }_{1}}h\left( x \right)\le 0.$ Consequently, at the boundary $ {{\left. \dot{h}\left( x \right) \right|}_{h\left( x \right)=0}}={{\left. {{{\tilde{h}}}_{1}}\left( x \right) \right|}_{h\left( x \right)=0}}={{\left( \dot{h}\left( x \right)-{{\lambda }_{1}}h\left( x \right) \right)}_{h\left( x \right)=0}}\le 0$,
  	and so, using~\eqref{eq65}, the set $H$ is also forward invariant with ${{\tilde{H}}_{1}}\subset H.$ For $r=2$, if ${{\tilde{H}}_{2}}$ is forward invariant, then
  	\begin{equation}
  		\label{eq67}
  		\begin{aligned} 
  			{{\tilde{h}}_{2}}\left( x \right)=\left( s-{{\lambda }_{2}} \right)\left( s-{{\lambda }_{1}} \right)h\left( x \right)=\left( s-{{\lambda }_{2}} \right){{\tilde{h}}_{1}}\left( x \right)\le 0
  		\end{aligned}
  	\end{equation}
  	implying ${{\tilde{H}}_{1}}$ is forward invariant, which in turn implies $H$ is forward invariant, ${{\tilde{H}}_{2}}\subset {{\tilde{H}}_{1}}\subset H.$
  	Continuation of this argument by induction, gives ${{\tilde{H}}_{r}}\subset {{\tilde{H}}_{r-1}}\subset \ldots \subset {{\tilde{H}}_{1}}\subset H.$
  \end{proof}

\begin{proof}[Proof of Lemma~\ref{lem1}]  
	For relative degree one limited outputs, 
  \begin{equation}
  	\label{eq62}
  	\begin{aligned} 
  		{{A}_{\text{CBF}}}=A-B{{\left( {{C}_{\lim }}B \right)}^{-1}}\left[ \begin{matrix}
  	{{C}_{\lim ,1}}\left( A+{{\alpha }_{1}}I \right) \\ 
  	{{C}_{\lim ,2}}\left( A+{{\alpha }_{2}}I \right) \\ 
  	\vdots  \\ 
  	{{C}_{\lim ,m}}\left( A+{{\alpha }_{m}}I \right) \\ 
  \end{matrix} \right]=A-B{{\left( {{C}_{\lim }}B \right)}^{-1}}\left( {{C}_{\lim }}A+\underbrace{\text{diag}\left( \begin{matrix}
  		{{\alpha }_{1}} & {{\alpha }_{2}} & \cdots  & {{\alpha }_{m}}  \\
  	\end{matrix} \right)}_{D}{{C}_{\lim }} \right)
  \end{aligned}
\end{equation}
  where $D=\text{diag}\left([ {{\alpha }_{1}} , {{\alpha }_{2}} , \cdots  , {{\alpha }_{m}}] \right)$. $P=B{{\left( {{C}_{\lim }}B \right)}^{-1}}{{C}_{\lim }}$ is an idempotent projection matrix with $P={{P}^{2}}$ and $\text{eig}(P)$ being either zero or one. Consider the coordinate transformation matrix and its inverse
  \begin{equation}
  	\label{eq63}
  	\begin{aligned} 
  		T=\left[ \begin{matrix}
  	B & {{C}_{\lim }}^{\bot }  \\
  \end{matrix} \right],{{T}^{-1}}=\left[ \begin{matrix}
  	{{\left( {{C}_{\lim }}B \right)}^{-1}}{{C}_{\lim }} \\ 
  	{{B}^{\bot }} \\ 
  \end{matrix} \right]
\end{aligned}
\end{equation}
  with ${{B}^{\bot }}$ and ${{C}_{\lim }}^{\bot }$ chosen such that ${{B}^{\bot }}B={{0}_{m\times m}},{{C}_{\lim }}{{C}_{\lim }}^{\bot }={{0}_{m\times m}},$ and ${{B}^{\bot }}{{C}_{\lim }}^{\bot }={{I}_{m}}$. Clearly, ${{T}^{-1}}T={{I}_{n}}$. Then, 
  \begin{equation}
  	\label{eq64}
  	\begin{aligned} 
  		{{T}^{-1}}{{A}_{\text{CBF}}}T
  =
   \begin{bmatrix}
  	-{{\left( {{C}_{\lim }}B \right)}^{-1}}D{{C}_{\lim }}B & 0  \\
  	{{B}^{\bot }}AB & {{B}^{\bot }}A{{C}^{\bot }} 
  \end{bmatrix} 
\end{aligned}
\end{equation}
  Hence, the eigenvalues of ${{A}_{\text{CBF}}}$ are given by $\text{eig}\left( {{B}^{\bot }}A{{C}_{\lim }}^{\bot } \right)$ and $\text{eig}(-{{\left( {{C}_{\lim }}B \right)}^{-1}}D{{C}_{\lim }}B)=\text{eig}(-D)$ with $-\alpha_{i}$ being the eigenvalues of the latter. All statements in Lemma~\ref{lem1} follow immediately as the plant eigenvalues and CBF eigenvalues are strictly disjoint (separated by the span and null space defining $P$).  
  \end{proof}

  \section*{Appendix B: Proofs of Theorems}
  
  \begin{proof}[Proof of Theorem~\ref{thm1}]  
Consider the Lagrangian of the QP in~\eqref{eq12} with the Lagrange multipliers $\gamma $
\begin{equation}
	\label{eq68}
	\begin{aligned} 
\mathcal{L}\left( x,\,\pi ,\gamma  \right)={{\pi }^{T}}{{R}_{\pi }}\pi +{{\left[ \begin{matrix}
			{{\gamma }_{\min}}  \\
			{{\gamma }_{\max}}  \\
		\end{matrix} \right]}^{T}}\left( \left[ \begin{matrix}
	-{{I}_{m}}  \\
	{{I}_{m}}  \\
\end{matrix} \right]{{H}_{\pi }}\pi +\Delta H\left( x \right) \right).
\end{aligned}
\end{equation}
Using~\eqref{eq68}, the KKT conditions for optimality \cite{BoydVandenberghe2004} are
\begin{equation}
	\label{eq69}
	\begin{aligned} 
	& \text{Stationarity}:\frac{\partial \mathcal{L}\left( x,\,\pi ,\gamma  \right)}{\partial \pi }=0 \\ 
	& \text{Primal }\,\text{Feasibility}:H\left( x,\,\pi  \right)\le 0 \\ 
	& \text{Dual}\,\text{ Feasibility}:\gamma \ge 0 \\ 
	& \text{Complementary}\,\text{ Slackness}:{{\gamma }_{i}}\,{{H}_{i}}\left( x,\,\pi  \right)=0\text{ for all }i. 
\end{aligned}
\end{equation}
KKT stationarity, $2\,{{R}_{\pi }}{{\pi }}-H_{\pi }^{T}\left( \gamma _{\min}^{\star }-\gamma _{\max}^{\star } \right)=0,$ rewritten as 
\begin{equation}
	\label{eq70}
	\begin{aligned} 
\pi =0.5R_{\pi }^{-1}H_{\pi }^{T}\left( \gamma _{\min}^{\star }-\gamma _{\max}^{\star } \right),
\end{aligned}
\end{equation}
yields the optimal policy. Next, inserting this optimal policy~\eqref{eq70} into the primal feasibility condition results in 
\begin{equation}
	\label{eq71}
	\begin{aligned} 
\left[ \begin{matrix}
	{{R}_{\gamma ,\gamma }} & -{{R}_{\gamma ,\gamma }}  \\
	-{{R}_{\gamma ,\gamma }} & {{R}_{\gamma ,\gamma }}  \\
\end{matrix} \right] \begin{bmatrix}
	\gamma _{\min}^{\star } \\ 
	 \gamma _{\max}^{\star } \\ 
\end{bmatrix} \ge 2 \begin{bmatrix}
	\Delta {{H}_{\min}}\left( x \right) \\ 
	\Delta {{H}_{\max}}\left( x \right) \\ 
\end{bmatrix} 
\end{aligned}
\end{equation}
with ${{R}_{\gamma ,\gamma }}={{H}_{\pi }}R_{\pi }^{-1}H_{\pi }^{T}.$ As ${{R}_{\gamma ,\gamma }}={{I}_{m}}$ with ${{R}_{\pi }}=H_{\pi }^{T}{{H}_{\pi }}$ is diagonal and the min/max bounds are mutually exclusive, i.e., ${{({{\gamma }_{\min}})}_{i}}{{({{\gamma }_{\max}})}_{i}}=0$ for all $i=1,...,m,$~\eqref{eq71} can be written as
$\gamma _{\min}^{\star }\ge 2\Delta {{H}_{\min}}\left( x \right),\text{     }\gamma _{\max}^{\star }\ge 2\Delta {{H}_{\max}}\left( x \right).$ 
Hence, any $({{\gamma }_{\min}},{{\gamma }_{\max}})$ solving $\gamma _{\min}^{\star }=\max \left( {{0}_{m\times 1}},2\Delta {{H}_{\min}}\left( x \right) \right),\text{ }\gamma _{\max}^{\star }=\max \left( {{0}_{m\times 1}},2\Delta {{H}_{\max}}\left( x \right) \right)$ satisfy primal feasibility and stationarity. 
Dual feasibility is guaranteed due to the max() operator, and the max() operator also ensures that the Lagrange multipliers are positive if, and only if, the corresponding constraint is active. Inserting the Lagrange multipliers into~\eqref{eq70} provides the optimal control augmentation $\pi ,$, and thus, proves optimality.
\end{proof}

\begin{proof}[Proof of Theorem~\ref{thm2}] 
This follows directly from Lemma~\ref{lem2} with the set ${{X}_{0}}=\{x:H(x,w)\le 0\}$. 
Boundedness can be shown using Bounded-Input Bounded-Output arguments leveraging global Lipschitz continuity of a switched LTI system and the fact that ${{y}_{\lim }}={{C}_{\lim }}x$ is bounded. Further, for switched linear systems, it is known that global uniform exponential stability (GUES) is guaranteed if and only if the system is locally attractive for every possible switching signal ${{\delta }_{i}}\left( x \right)$ \cite{Liberzon2003}. Thus, results from the literature on switched systems such as dwell time can be used \cite{Liberzon2003}. In particular, the proposed control solution can be analyzed as switched state feedback system, and thus, Theorem 6 in \cite{Hespanha2004} is directly applicable. GUES can be shown if the operational boundaries are sufficiently far apart, or the dynamics are sufficiently fast such that frequent switching between active constraints is prevented.
\end{proof}

  \begin{proof}[Proof of Theorem~\ref{thm4}]  
  	For the PI servo-control design, the analytical solution uses
  \begin{equation}
  	\label{eq86}
  	\begin{aligned} 
  	H_{\pi }^{-1}=\left[ \begin{matrix}
  	-K_{I}^{-1} & K_{I}^{-1}\left( -{{K}_{I}}{{D}_{p\,\,\text{reg}}}-{{K}_{P}}{{B}_{p}} \right)H_{u}^{-1}  \\
  	{{0}_{m\times m}} & H_{u}^{-1}  \\
  \end{matrix} \right]
\end{aligned}
\end{equation}
  with ${{H}_{\pi }}$ being non-singular. Eq.~\eqref{eq86} shows that the control augmentation ${{w}^{\star }}$ is only used for output constraints. Hence, if ${{z}_{\lim }}$ is within its min/max bounds but a component ${{({{u}_{bl}})}_{i}}$ of the baseline signal reaches its min or max value, then $w=0$ and $v\ne 0$ will enforce the limits. Next, consider the case of the limited output ${{({{z}_{\lim }})}_{i}}$ reaching a bound. If under baseline control, $\left| {{({{z}_{\lim }})}_{i}} \right|>z_{\lim }^{\max }$ then $w$ must be found to satisfy the output constraint. Due to the system’s linearity, the output depends linearly on the total control input. Therefore, the absolute value of the total control input must be reduced, which in turn implies that $w,$ as an incremental signal, must act in the opposite direction of the baseline control input, ${{({{u}_{\rm bl}})}_{i}}{{w}_{i}}\le 0$, for all components. Then,
  \begin{equation}
  	\label{eq87}
  	\begin{aligned} 
  	 \left( {{u}_{\rm bl}}+w \right)_{i}^{2}=\left( {{u}_{\rm bl}} \right)_{i}^{2}-2\left| {{\left( {{u}_{\rm bl}} \right)}_{i}}{{w}_{i}} \right|+w_{i}^{2}  ={{\left( \left| {{\left( {{u}_{\rm bl}} \right)}_{i}} \right|-\left| {{w}_{i}} \right| \right)}^{2}}\le {{\left| {{\left( {{u}_{\rm bl}} \right)}_{i}} \right|}^{2}}\le \left( u_{\rm bl}^{\max } \right)_{i}^{2}  
\end{aligned}
\end{equation}
  and consequently, $\left| {{({{u}_{\rm bl}}+w)}_{i}} \right|\le {{(u_{\rm bl}^{\max })}_{i}}.$
  UUB of ${{e}_{yI}}=-K_{I}^{-1}({{u}_{\rm bl}}-{{K}_{p}}{{x}_{p}})$ is given directly by boundedness of ${{u}_{\rm bl}}$ (proved in  Theorem~\ref{thm1}) and boundedness of ${{x}_{p}}$ (proved in Theorem~\ref{thm1}).
    \end{proof}

  \begin{proof}[Proof of Theorem~\ref{thm3}]
  	By construction, for any relative degree ${{r}_{i}}$ of constraint $i$, the CBF augmentation enforces
  	\begin{equation}
  		\label{eq73}
  		\begin{aligned} 
  			\left[ \begin{matrix}
  				-{{C}_{\lim ,i}}B  \\
  				{{C}_{\lim ,i}}B  \\
  			\end{matrix} \right]\pi \left( t \right)+\left[ \begin{matrix}
  				-{{C}_{\lim ,i}}\left( A+\alpha _{i}^{\star }{{I}_{n}} \right)\hat{x}\left( t \right)-{{C}_{\lim ,i}}B{{u}_{\text{bl}}}\left( t \right)+\alpha _{i}^{\star }y_{\lim ,i}^{\min }  \\
  				{{C}_{\lim ,i}}\left( A+\alpha _{i}^{\star }{{I}_{n}} \right)\hat{x}\left( t \right)+{{C}_{\lim ,i}}B{{u}_{\text{bl}}}\left( t \right)-\alpha _{i}^{\star }y_{\lim ,i}^{\max }  \\
  			\end{matrix} \right]\le 0
  		\end{aligned}
  	\end{equation}
  	i.e., for relative degrees greater than one, ${{C}_{\lim ,i}}B=0$ but~\eqref{eq73} still holds. This is true since the modified constraints in~\eqref{eq10} are a more conservative inequality used for control augmentation design, i.e., any control augmentation signal designed using the modified constraints also satisfies~\eqref{eq73}. 
  	Replacing the state estimate $\hat x(t) = x\left( t \right)-{{e}_{\text{est}}}\left( t \right)$ and rewriting the modified constraint and the state $x$, 
  	\begin{equation}
  		\label{eq75}
  		\begin{aligned} 
  			\left[ \begin{matrix}
  				\left( s+\alpha _{i}^{\star } \right){{h}_{\min ,i}}\left( x\left( t \right) \right)  \\
  				\left( s+\alpha _{i}^{\star } \right){{h}_{\max ,i}}\left( x\left( t \right) \right)  \\
  			\end{matrix} \right]\le \left[ \begin{matrix}
  				-{{C}_{\lim ,i}}\left( A+\alpha _{i}^{\star }{{I}_{n}} \right){{e}_{\text{est}}}\left( t \right)  \\
  				{{C}_{\lim ,i}}\left( A+\alpha _{i}^{\star }{{I}_{n}} \right){{e}_{\text{est}}}\left( t \right)  \\
  			\end{matrix} \right]
  		\end{aligned}
  	\end{equation}
  	Next, consider the min constraint in~\eqref{eq75},
  	\begin{equation}
  		\label{eq76}
  		\begin{aligned} 
  			\left( s+\alpha _{i}^{\star } \right){{h}_{\min ,i}}\left( x\left( t \right) \right)={{\dot{h}}_{\min ,i}}\left( x\left( t \right) \right)+\alpha _{i}^{\star }{{h}_{\min ,i}}\left( x\left( t \right) \right)\le -{{C}_{\lim ,i}}\left( A+\alpha _{i}^{\star }{{I}_{n}} \right){{e}_{\text{est}}}\left( t \right)
  		\end{aligned}
  	\end{equation}
  	We use the method of integrating factors and a series of conservative bounds. We multiply~\eqref{eq76} by ${{e}^{\alpha _{i}^{\star }t}}>0$, 
  	\begin{equation}
  		\label{eq77} 
  		\begin{matrix} 
  			{{e}^{\alpha _{i}^{\star }t}}\left( {{{\dot{h}}}_{\min ,i}}\left( x\left( t \right) \right)+\alpha _{i}^{\star }{{h}_{\min ,i}}\left( x\left( t \right) \right) \right)
  			= {{e}^{\alpha _{i}^{\star }t}}\frac{\text{d}{{h}_{\min ,i}}\left( x\left( t \right) \right)}{\text{d}t}+\frac{\text{d}{{e}^{\alpha _{i}^{\star }t}}}{\text{d}t}{{h}_{\min ,i}}\left( x\left( t \right) \right)
  			\le 
  			-{{e}^{\alpha _{i}^{\star }t}}{{C}_{\lim ,i}}\left( A+\alpha _{i}^{\star }{{I}_{n}} \right){{e}_{\text{est}}}\left( t \right)
  			\\
  			\frac{\text{d}}{\text{d}t}\left( {{e}^{\alpha _{i}^{\star }t}}{{h}_{\min ,i}}\left( x\left( t \right) \right) \right)\le -{{e}^{\alpha _{i}^{\star }t}}{{C}_{\lim ,i}}\left( A+\alpha _{i}^{\star }{{I}_{n}} \right){{e}_{\text{est}}}\left( t \right).
  		\end{matrix}
  	\end{equation}
  	Since~\eqref{eq77} holds (in a differentiated sense), it also holds in an integral sense,
  	\begin{equation}
  		\label{eq79}
  		\begin{aligned} 
  			{{e}^{\alpha _{i}^{\star }t}}{{h}_{\min ,i}}\left( x\left( t \right) \right)\le {{h}_{\min ,i}}\left( x\left( t=0 \right) \right)+\int{-{{C}_{\lim ,i}}\left( A+\alpha _{i}^{\star }{{I}_{n}} \right){{e}_{\text{est}}}\left( t \right){{e}^{\alpha _{i}^{\star }\tau }}\text{d}\tau }.
  		\end{aligned}
  	\end{equation}
  	The state estimation error ${{e}_{\text{est}}}\left( t \right)$ decays exponentially and there exists a $k$ such that $C_{\lim ,i}\left( A+\alpha _{i}^{\star }{{I}_{n}} \right){{e}_{\text{est}}}\left( t \right)\le k\| {{e}_{\text{est}}}\left( t=0 \right) \|{{e}^{{{\lambda }_{\max }}t}}$ for all $t\ge 0$ with the maximum (slowest) eigenvalue ${{\lambda }_{\max }}$. Note that $k$ can be explicitly stated using the system parameters. Thus,
  	\begin{equation}
  		\label{eq80}
  		\begin{aligned} 
  			{{e}^{\alpha _{i}^{\star }t}}{{h}_{\min ,i}}\left( x\left( t \right) \right)\le {{h}_{\min ,i}}\left( x\left( t=0 \right) \right)+\int{k\left\| {{e}_{\text{est}}}\left( t=0 \right) \right\|{{e}^{{{\lambda }_{\max }}\tau }}{{e}^{\alpha _{i}^{\star }\tau }}\text{d}\tau }.
  		\end{aligned}
  	\end{equation}
  	Integrating and dividing by ${{e}^{{{\alpha }_{0i}}t}}>0$ yields
  	\begin{equation}
  		\label{eq81}
  		\begin{aligned}  
  			{{h}_{\min ,i}}\left( x\left( t \right) \right)\le {{e}^{-\alpha _{i}^{\star }t}}{{h}_{\min ,i}}\left( x\left( t=0 \right) \right)+\frac{k\left\| {{e}_{\text{est}}}\left( t=0 \right) \right\|}{\left| {{\lambda }_{\max }}+\alpha _{i}^{\star } \right|}{{e}^{{{\lambda }_{\max }}t}}. 
  		\end{aligned}
  	\end{equation}
  	Finally, it is clear that a sufficient (conservative) condition for ${{h}_{\min ,i}}\left( x\left( t \right) \right)\le 0$ is 
  	\begin{equation}
  		\label{eq82}
  		\begin{aligned}  
  			{{h}_{\min ,i}}\left( x\left( t \right) \right)\le {{e}^{-\alpha _{i}^{\star }t}}{{h}_{\min ,i}}\left( x\left( t=0 \right) \right)+\frac{k\left\| {{e}_{\text{est}}}\left( t=0 \right) \right\|}{\left| {{\lambda }_{\max }}+\alpha _{i}^{\star } \right|}{{e}^{{{\lambda }_{\max }}t}}\le 0.
  		\end{aligned}
  	\end{equation}
  	Rearranging the terms in~\eqref{eq82} yields the following conservative condition for constraint satisfaction
  	\begin{equation}
  		\label{eq83}
  		\begin{aligned} 
  			{{e}^{\left( {{\lambda }_{\max }}+\alpha _{i}^{\star } \right)t}}\le -\frac{\left| {{\lambda }_{\max }}+\alpha _{i}^{\star } \right|}{k\left\| {{e}_{\text{est}}}\left( t=0 \right) \right\|}{{h}_{\min ,i}}\left( x\left( t=0 \right) \right).
  		\end{aligned}
  	\end{equation}
  	Clearly, for~\eqref{eq83} to hold forward in time $t>{{t}_{h\le 0}}>0$, $\left( {{\lambda }_{\max }}+{{\alpha }_{0i}} \right)<0$ to keep the left hand side bounded. 
  	Finally, we can explicitly compute the time
  	\begin{equation}
  		\label{eq84}
  		\begin{aligned} 
  			t\ge {{t}_{h\le 0}}=\frac{1}{\left( {{\lambda }_{\max }}+\alpha _{i}^{\star } \right)}\ln \left( -\frac{\left| {{\lambda }_{\max }}+\alpha _{i}^{\star } \right|}{k\left\| {{e}_{\text{est}}}\left( t=0 \right) \right\|}{{h}_{\min ,i}}\left( x\left( t=0 \right) \right) \right).
  		\end{aligned}
  	\end{equation}
  	It can also be seen that for small initial state estimation errors ${{e}_{\text{est}}}\left( t=0 \right)$, the logarithmic term can become positive, 
  	implying constraint satisfaction is guaranteed forward in time with ${{t}_{h\le 0}}=0$.
  \end{proof}

\bibliography{mybibfileExport}

@inproceedings{fisher2025analytical,
  title={Analytical construction of cbf-based safety filters for simultaneous state and input constraints},
  author={Fisher, Peter A and Annaswamy, Anuradha M},
  booktitle={2025 IEEE 64th Conference on Decision and Control (CDC)},
  pages={6006--6012},
  year={2025},
  organization={IEEE}
}

@inproceedings{cohen2026control,
  title={Control barrier functions for flight-critical systems in aerospace applications},
  author={Cohen, Max H and Hussain, Heather and Molnar, Tamas G and Van Wijk, David EJ and Gaudio, Joseph and Fisher, Peter A and Menner, Marcel and Annaswamy, Anuradha and Autenrieb, Johannes and Lavretsky, Eugene and others},
  booktitle={2026 American Control Conference (ACC)},
  pages={49--66},
  year={2026},
  organization={IEEE}
}

@book{Lavretsky2024,
  author = {E. Lavretsky and K. A. Wise},
  title = {Robust adaptive control},
  subtitle = {Robust and adaptive control: With aerospace applications, Second Edition, Advanced Textbooks in Control and Signal Processing},
  publisher = {Springer Nature Switzerland AG},
  year = {2024}
}

@book{Hespanha2018,
  author = {J. Hespanha},
  title = {Linear systems theory},
  publisher = {Princeton university press},
  year = {2018}
}

@book{AstromMurray2021,
  author = {K. Åström and R. Murray},
  title = {Feedback systems: an introduction for scientists and engineers},
  address = {Princeton, NJ},
  publisher = {Princeton university press},
  year = {2021}
}

@inproceedings{LavretskyMenner2025,
  author = {E. Lavretsky and M. Menner},
  title = {Servo-Controllers for Linear Time-Invariant Systems with Operational Constraints},
  booktitle = {American Control Conference (ACC)},
  year = {2025}
}

@inproceedings{MennerLavretsky2026,
  author = {M. Menner and E. Lavretsky},
  title = {Robust Linear Design for Flight Control Systems with Operational Constraints},
  booktitle = {AIAA SCITECH 2026 Forum},
  year = {2026}
}

@article{Nagumo1942,
  author = {M. Nagumo},
  title = {Über die Lage der Integralkurven gewöhnlicher Differentialgleichungen},
  journal = {Proceedings of the Physico-Mathematical Society of Japan. 3rd Series},
  volume = {24},
  pages = {551-559},
  year = {1942}
}

@article{MennerLavretsky2024,
  author = {M. Menner and E. Lavretsky},
  title = {Translation of Nagumo's Foundational Work on Barrier Functions: On the Location of Integral Curves of Ordinary Differential Equations},
  journal = {arXiv preprint arXiv:2406.18614},
  year = {2024}
}

@article{McNabb1986,
  author = {A. McNabb},
  title = {Comparison theorems for differential equations},
  journal = {Journal of mathematical analysis and applications},
  volume = {119},
  number = {1-2},
  pages = {417-428},
  year = {1986}
}

@article{FreemanKokotovic1996,
  author = {R. Freeman and P. Kokotovic},
  title = {Inverse optimality in robust stabilization},
  journal = {SIAM journal on control and optimization},
  volume = {34},
  number = {4},
  pages = {1365-1391},
  year = {1996}
}

@book{BoydVandenberghe2004,
  author = {S. Boyd and L. Vandenberghe},
  title = {Convex optimization},
  address = {Cambridge, MA},
  publisher = {Cambridge university press},
  year = {2004}
}

@article{AmesXuGrizzleTabuada2016,
  author = {A. D. Ames and X. Xu and J. W. Grizzle and P. Tabuada},
  title = {Control barrier function based quadratic programs for safety critical systems},
  journal = {IEEE Trans. Automatic Control},
  volume = {62},
  number = {8},
  pages = {3861-3876},
  year = {2016}
}

@article{WabersichZeilinger2022,
  author = {K. Wabersich and M. Zeilinger},
  title = {Predictive control barrier functions: Enhanced safety mechanisms for learning-based control},
  journal = {IEEE Trans. Automatic Control},
  volume = {68},
  number = {5},
  pages = {2638-2651},
  year = {2022}
}

@article{WabersichEtAl2023,
  author = {K. Wabersich and A. Taylor and J. Choi and K. Sreenath and C. Tomlin and A. Ames and M. Zeilinger},
  title = {Data-driven safety filters: Hamilton-jacobi reachability, control barrier functions, and predictive methods for uncertain systems},
  journal = {IEEE Control Systems Magazine},
  volume = {43},
  number = {5},
  pages = {137-177},
  year = {2023}
}

@article{MorariLee1999,
  author = {M. Morari and J. Lee},
  title = {Model predictive control: past, present and future},
  journal = {Computers \& chemical engineering},
  volume = {23},
  number = {4-5},
  pages = {667-682},
  year = {1999}
}

@article{SeoLeeBaekHorowitzChoi2022,
  author = {J. Seo and J. Lee and E. Baek and R. Horowitz and J. Choi},
  title = {Safety-critical control with nonaffine control inputs via a relaxed control barrier function for an autonomous vehicle},
  journal = {IEEE Robotics and Automation Letters},
  volume = {7},
  number = {2},
  pages = {1944-1951},
  year = {2022}
}

@article{XiaoWangHasaniChahineAminiLiRus2023,
  author = {W. Xiao and T. Wang and R. Hasani and M. Chahine and A. Amini and X. Li and D. Rus},
  title = {Barriernet: Differentiable control barrier functions for learning of safe robot control},
  journal = {IEEE Trans. Robotics},
  volume = {39},
  number = {3},
  pages = {2289-2307},
  year = {2023}
}

@inproceedings{AmesCooganEgerstedtNotomistaSreenathTabuada2019,
  author = {A. D. Ames and S. Coogan and M. Egerstedt and G. Notomista and K. Sreenath and P. Tabuada},
  title = {Control barrier functions: Theory and applications},
  booktitle = {European Control Conference},
  pages = {3420-3431},
  year = {2019}
}

@article{Blanchini1999,
  author = {F. Blanchini},
  title = {Set invariance in control},
  journal = {Automatica},
  volume = {35},
  number = {11},
  pages = {1747-1767},
  year = {1999}
}

@inproceedings{TaylorAmes2020,
  author = {A. Taylor and A. Ames},
  title = {Adaptive safety with control barrier functions},
  booktitle = {American Control Conference},
  pages = {1399-1405},
  year = {2020}
}

@article{XiaoBelta2021,
  author = {W. Xiao and C. Belta},
  title = {High-order control barrier functions},
  journal = {IEEE Trans. Automatic Control},
  volume = {67},
  number = {7},
  pages = {3655-3662},
  year = {2021}
}

@inproceedings{NguyenSreenath2016,
  author = {Q. Nguyen and K. Sreenath},
  title = {Exponential control barrier functions for enforcing high relative-degree safety-critical constraints},
  booktitle = {2016 American Control Conference (ACC)},
  pages = {322-328},
  year = {2016}
}

@article{MolnarKissAmesOrosz2022,
  author = {T. Molnar and A. Kiss and A. Ames and G. Orosz},
  title = {Safety-critical control with input delay in dynamic environment},
  journal = {IEEE Trans. Control Systems Technology},
  volume = {31},
  number = {4},
  pages = {1507-1520},
  year = {2022}
}

@article{AllibhoyCortes2023,
  author = {A. Allibhoy and J. Cortés},
  title = {Control barrier function-based design of gradient flows for constrained nonlinear programming},
  journal = {IEEE Trans. Automatic Control},
  year = {2023}
}

@book{Liberzon2003,
  author = {D. Liberzon},
  title = {Switching in systems and control},
  volume = {190},
  address = {Boston, MA},
  publisher = {Birkhauser},
  year = {2003}
}

@article{Hespanha2004,
  author = {J. Hespanha},
  title = {Uniform stability of switched linear systems: Extensions of LaSalle's invariance principle},
  journal = {IEEE Trans. Automatic Control},
  volume = {49},
  number = {4},
  pages = {470-482},
  year = {2004}
}

@article{Luenberger1964,
  author = {D. Luenberger},
  title = {Observing the State of a Linear System},
  journal = {IEEE Transactions on Military Electronics},
  volume = {8},
  number = {2},
  pages = {74-80},
  year = {1964}
}

@article{Luenberger1966,
  author = {D. Luenberger},
  title = {Observers for multivariable systems},
  journal = {IEEE Transactions on Automatic Control},
  volume = {11},
  number = {2},
  pages = {190-197},
  year = {1966}
}

@book{Khalil2001,
  author = {H. Khalil},
  title = {Nonlinear Systems},
  edition = {3rd Edition},
  address = {East Lansing, MI},
  publisher = {Pearson},
  year = {2001}
}

@article{AlanTaylorHeAmesOrosz2023,
  author = {A. Alan and A. Taylor and C. He and A. Ames and G. Orosz},
  title = {Control barrier functions and input-to-state safety with application to automated vehicles},
  journal = {IEEE Trans. Control Systems Technology},
  volume = {31},
  number = {6},
  pages = {2744-2759},
  year = {2023}
}

@article{menner2026output,
  title={Output Feedback Control of Linear Time-Invariant Systems with Operational Constraints},
  author={Menner, Marcel and Hussain, Heather and Lavretsky, Eugene},
  journal={arXiv preprint arXiv:2605.21399},
  year={2026}
}

\end{document}